\input harvmac.tex


\input epsf.tex
\def\figin{\epsfcheck\figin}\def\figins{\epsfcheck\figins}
\def\epsfcheck{\ifx\epsfbox\UnDeFiNeD
\message{(NO epsf.tex, FIGURES WILL BE IGNORED)}
\gdef\figin##1{\vskip2in}\gdef\figins##1{\hskip.5in}
\else\message{(FIGURES WILL BE INCLUDED)}%
\gdef\figin##1{##1}\gdef\figins##1{##1}\fi}
\def\DefWarn#1{}
\def\figinsert{\goodbreak\midinsert}
\def\ifig#1#2#3{\DefWarn#1\xdef#1{fig.~\the\figno}
\writedef{#1\leftbracket fig.\noexpand~\the\figno}%
\figinsert\figin{\centerline{#3}}\medskip\centerline{\vbox{\baselineskip12pt
\advance\hsize by -1truein\noindent\footnotefont{\bf
Fig.~\the\figno:} #2}}
\bigskip\endinsert\global\advance\figno by1}

\def\ndt{\noindent}

\def\Ga{{\Gamma}}

\def\K3{{\bf K3}}
\def\journal#1&#2(#3){\unskip, \sl #1\ \bf #2 \rm(19#3) }
\def\andjournal#1&#2(#3){\sl #1~\bf #2 \rm (19#3) }

\def\tilde{\widetilde}

\def\frac#1#2{{#1\over#2}}

\def\vev#1{\langle#1\rangle}

\def\inbar{\,\vrule height1.5ex width.4pt depth0pt}
\def\IC{\relax\hbox{$\inbar\kern-.3em{\rm C}$}}
\def\IR{\relax{\rm I\kern-.18em R}}
\def\IP{\relax{\rm I\kern-.18em P}}

%
%


%
\catcode`\@=11
\def\slash#1{\mathord{\mathpalette\c@ncel{#1}}}
\overfullrule=0pt

\def\CC{{\cal C}}

\def\FF{{\cal F}}

\def\NN{{\cal N}}

\def\QQ{{\cal Q}}

\def\SS{{\cal S}}

\def\ZZ{{\cal Z}}
\def\lam{\lambda}

\def\underrel#1\over#2{\mathrel{\mathop{\kern\z@#1}\limits_{#2}}}

\catcode`\@=12
\def\rt{ \rho_1}

%

\def\vev#1{\left\langle #1 \right\rangle}

\def\tr{{\rm tr}}

\def\exp{{\rm exp}}


\def \ov {\over}
\def \p {\partial}
\def \ha {{1 \ov 2}}

\def \lam {\lambda}

\def \om {\omega}
\def \Om {\Omega}
\def \ep {\epsilon}

\def \ga {\gamma}

\def\le{\left}
\def\ri{\right}

\def\th{\theta}

\def\IL{\relax{\rm I\kern-.18em L}}
\def\IH{\relax{\rm I\kern-.18em H}}
\def\IR{\relax{\rm I\kern-.18em R}}
\def\IC{\relax\hbox{$\inbar\kern-.3em{\rm C}$}}





\def\makeblankbox#1#2{\hbox{\lower\dp0\vbox{\hidehrule{#1}{#2}%
\kern -#1
\hbox to \wd0{\hidevrule{#1}{#2}%
\raise\ht0\vbox to #1{}
\lower\dp0\vtop to #1{}
\hfil\hidevrule{#2}{#1}}%
\kern-#1\hidehrule{#2}{#1}}}%
}%
\def\hidehrule#1#2{\kern-#1\hrule height#1 depth#2 \kern-#2}%
\def\hidevrule#1#2{\kern-#1{\dimen0=#1\advance\dimen0 by #2\vrule
width\dimen0}\kern-#2}%
\def\openbox{\ht0=1.2mm \dp0=1.2mm \wd0=2.4mm \raise 2.75pt
\makeblankbox {.25pt} {.25pt} }

\def\bun#1/#2{\leavevmode
\kern.1em \raise .5ex \hbox{\the\scriptfont0 #1}%
\kern-.1em $/$%
\kern-.15em \lower .25ex \hbox{\the\scriptfont0 #2}%
}

\def\opensquare{\ht0=3.4mm \dp0=3.4mm \wd0=6.8mm \raise 2.7pt
\makeblankbox {.25pt} {.25pt} }


\def\sector#1#2{\ {\scriptstyle #1}\hskip 1mm
\mathop{\opensquare}\limits_{\lower
1mm\hbox{$\scriptstyle#2$}}\hskip 1mm}

\def\tsector#1#2{\ {\scriptstyle #1}\hskip 1mm
\mathop{\opensquare}\limits_{\lower
1mm\hbox{$\scriptstyle#2$}}^\sim\hskip 1mm}


\lref\LV{Landsman and Van Weert , Phys.\ Rept.\ }

\lref\CrnkovicMS{ C.~Crnkovic, M.~R.~Douglas and G.~W.~Moore,
``Physical Solutions For Unitary Matrix Models,'' Nucl.\ Phys.\ B
{\bf 360}, 507 (1991).
}

\lref\SchnitzerQT{ H.~J.~Schnitzer, ``Confinement / deconfinement
transition of large N gauge theories with N(f) fundamentals:
N(f)/N finite,'' hep-th/0402219.
}

\lref\aharony{ O.~Aharony, J.~Marsano, S.~Minwalla, K.~Papadodimas
and M.~Van Raamsdonk, ``The Hagedorn / deconfinement phase
transition in weakly coupled large N gauge theories,''
hep-th/0310285.
}

\lref\sundborg{ B.~Sundborg, ``The Hagedorn transition,
deconfinement and N = 4 SYM theory,'' Nucl.\ Phys.\ B {\bf 573},
349 (2000) [hep-th/9908001];
B.~Sundborg, ``Stringy gravity, interacting tensionless strings
and massless higher spins,'' Nucl.\ Phys.\ Proc.\ Suppl.\ {\bf
102}, 113 (2001) [hep-th/0103247].
}

\lref\gw{ D.~J.~Gross and E.~Witten, ``Possible Third Order Phase
Transition In The Large N Lattice Gauge Theory,'' Phys.\ Rev.\ D
{\bf 21}, 446 (1980).
}

\lref\wadiapre{ S.~Wadia, ``A Study Of U(N) Lattice Gauge Theory
In Two-Dimensions,'' EFI-79/44-CHICAGO
}

\lref\wadia{ S.~R.~Wadia, ``N = Infinity Phase Transition In A
Class Of Exactly Soluble Model Lattice Gauge Theories,'' Phys.\
Lett.\ B {\bf 93}, 403 (1980).
}

\lref\malda{ J.~M.~Maldacena, ``The large N limit of
superconformal field theories and supergravity,'' Adv.\ Theor.\
Math.\ Phys.\ {\bf 2}, 231 (1998) [Int.\ J.\ Theor.\ Phys.\ {\bf
38}, 1113 (1999)] [hep-th/9711200].
}

\lref\malmag{O. Aharony, S.S. Gubser, J. Maldacena, H. Ooguri, Y.
Oz, ``Large N Field Theories, String Theory and Gravity," Phys.
Rep. {\bf 323} (2000) 183 [arXiv:hep-th/9905111].
}

\lref\davidrep{J. R. David, G. Mandal, S. R. Wadia, ``Microscopic
Formulation of Black Holes in String Theory," Phys.Rept. {\bf 369}
(2002) 549-686 [arXiv:hep-th/0203048].
}

\lref\witten{ E.~Witten, ``Anti-de Sitter space, thermal phase
transition, and confinement in gauge theories,'' Adv.\ Theor.\
Math.\ Phys.\ {\bf 2}, 505 (1998) [arXiv:hep-th/9803131].
}

\lref\hawkingpage{ S.~W.~Hawking and D.~N.~Page, ``Thermodynamics
Of Black Holes In Anti-De Sitter Space,'' Commun.\ Math.\ Phys.\
{\bf 87}, 577 (1983).
}

\lref\hawkingmoss{ S.W. Hawking, I.G. Moss, ``Supercooled Phase
Transition in the Very Early Universe", Phys.Lett.{\bf
B110}:35,1982. }

\lref\arnold{V.I. Arnold, ``Mathematical Methos of Classical
Mechanics", Springer-Verlag. }

\lref\callancoleman{C.G. Callan and S. Coleman, ``The fate of the
False Vacuum.2. First Quantum Corrections" Phys.Rev.{\bf
D16}:1762-1768,1977. }

\lref\langer{J.S. Langer, ``Theory of the Condensation Point",
Annals Phys.{\bf 41}:108-157,1967, Annals Phys.{\bf
281}:941-990,2000 }

\lref\mathur{Oleg Lunin, Samir D. Mathur, ``AdS/CFT duality and
the black hole information paradox", Nucl.Phys. B623 (2002)
342-394 [arXiv:hep-th/0109154]
}

\lref\atic{ J.~J.~Atick and E.~Witten, ``The Hagedorn Transition
And The Number Of Degrees Of Freedom Of String Theory,'' Nucl.\
Phys.\ B {\bf 310}, 291 (1988).
}

\lref\sapa{ B.~Sathiapalan, ``Vortices On The String World Sheet
And Constraints On Toral Compactification,'' Phys.\ Rev.\ D {\bf
35}, 3277 (1987).
}

\lref\kogan{ Y.~I.~Kogan, ``Vortices On The World Sheet And
String's Critical Dynamics,'' JETP Lett.\ {\bf 45}, 709 (1987)
[Pisma Zh.\ Eksp.\ Teor.\ Fiz.\ {\bf 45}, 556 (1987)].
}

\lref\goldsm{ Y.~Y.~Goldschmidt, ``1/N Expansion In
Two-Dimensional Lattice Gauge Theory,'' J.\ Math.\ Phys.\ {\bf
21}, 1842 (1980).
}

\lref\polaP{ J.~Jurkiewicz and K.~Zalewski, ``Vacuum Structure Of
The U(N $\to$ Infinity) Gauge Theory On A Two-Dimensional Lattice
For A Broad Class Of Variant Actions,'' Nucl.\ Phys.\ B {\bf 220},
167 (1983).
}

\lref\periw{ V.~Periwal and D.~Shevitz, ``Unitary Matrix Models As
Exactly Solvable String Theories,'' Phys.\ Rev.\ Lett.\ {\bf 64},
1326 (1990); \quad

V.~Periwal and D.~Shevitz, ``Exactly Solvable Unitary Matrix
Models: Multicritical Potentials And
Nucl.\ Phys.\ B {\bf 344}, 731 (1990).
}

\lref\kms{ I.~R.~Klebanov, J.~Maldacena and N.~Seiberg, ``Unitary
and complex matrix models as 1-d type 0 strings,'' hep-th/0309168.
}

\lref\klebanov{ I.~R.~Klebanov, ``Touching random surfaces and
Liouville gravity,'' Phys.\ Rev.\ D {\bf 51}, 1836 (1995)
[hep-th/9407167].
}

\lref\klebC{ I.~R.~Klebanov and A.~Hashimoto, ``Nonperturbative
solution of matrix models modified by trace squared terms,''
Nucl.\ Phys.\ B {\bf 434}, 264 (1995) [hep-th/9409064]; \quad

J.~L.~F.~Barbon, K.~Demeterfi, I.~R.~Klebanov and C.~Schmidhuber,
``Correlation functions in matrix models modified by wormhole
terms,'' Nucl.\ Phys.\ B {\bf 440}, 189 (1995) [hep-th/9501058].
}

\lref\das{ S.~R.~Das, A.~Dhar, A.~M.~Sengupta and S.~R.~Wadia,
``New Critical Behavior In D = 0 Large N Matrix Models,'' Mod.\
Phys.\ Lett.\ A {\bf 5}, 1041 (1990).
}

\lref\BranV{ R.~H.~Brandenberger and C.~Vafa, ``Superstrings In
The Early Universe,'' Nucl.\ Phys.\ B {\bf 316}, 391 (1989).
}

\lref\TanI{ N.~Deo, S.~Jain, O.~Narayan and C.~I.~Tan, ``The
Effect of topology on the thermodynamic limit for a string gas,''
Phys.\ Rev.\ D {\bf 45}, 3641 (1992).
}

\lref\polyakov{ A.~M.~Polyakov, ``Gauge fields and space-time,''
Int.\ J.\ Mod.\ Phys.\ A {\bf 17S1}, 119 (2002)
[arXiv:hep-th/0110196].
}

\lref\bipz{E.~Brezin, C.~Itzykson, G.~Parisi and J.~B.~Zuber,
``Planar Diagrams,'' Commun.\ Math.\ Phys.\ {\bf 59}, 35 (1978).
}

\lref\HallinKM{ J.~Hallin and D.~Persson, ``Thermal phase
transition in weakly interacting, large N(c) {QCD},'' Phys.\
Lett.\ B {\bf 429}, 232 (1998) [arXiv:hep-ph/9803234].
}

\lref\FuruuchiSY{ K.~Furuuchi, E.~Schreiber and G.~W.~Semenoff,
``Five-brane thermodynamics from the matrix model,''
arXiv:hep-th/0310286;
\quad G.~W.~Semenoff, ``Matrix model thermodynamics,''
arXiv:hep-th/0405107.
}

\lref\GubserBC{ S.~S.~Gubser, I.~R.~Klebanov and A.~M.~Polyakov,
``Gauge theory correlators from non-critical string theory,''
Phys.\ Lett.\ B {\bf 428}, 105 (1998) [arXiv:hep-th/9802109].
}

\lref\minW{ O.~Aharony, J.~Marsano, S.~Minwalla, K.~Papadodimas
and M.~Van Raamsdonk, ``A first order deconfinement transition in
large N Yang-Mills theory on a small 3-sphere,'' hep-th/0502149.
}

\lref\pisarski{ A.~Dumitru, J.~Lenaghan and R.~D.~Pisarski,
``Deconfinement in matrix models about the Gross-Witten point,''
hep-ph/0410294; \quad A.~Dumitru, Y.~Hatta, J.~Lenaghan,
K.~Orginos and R.~D.~Pisarski, ``Deconfining phase transition as a
matrix model of renormalized Polyakov loops,'' Phys.\ Rev.\ D {\bf
70}, 034511 (2004), hep-th/0311223.
}

\lref\WittenQJ{ E.~Witten, ``Anti-de Sitter space and
holography,'' Adv.\ Theor.\ Math.\ Phys.\ {\bf 2}, 253 (1998)
[arXiv:hep-th/9802150].
}

\lref\AharonyIG{ O.~Aharony, J.~Marsano, S.~Minwalla and
T.~Wiseman, ``Black hole - black string phase transitions in
thermal 1+1 dimensional
arXiv:hep-th/0406210.
}

\lref\FidkowskiFC{ L.~Fidkowski and S.~Shenker, ``D-brane
instability as a large N phase transition,'' arXiv:hep-th/0406086.
}

\lref\HaggiManiRU{ P.~Haggi-Mani and B.~Sundborg, ``Free large N
supersymmetric Yang-Mills theory as a string theory,'' JHEP {\bf
0004}, 031 (2000) [arXiv:hep-th/0002189].
}

\lref\malet{ J.~M.~Maldacena, ``Eternal black holes in
Anti-de-Sitter,'' JHEP {\bf 0304}, 021 (2003)
[arXiv:hep-th/0106112].
}

\lref\maldabarbon{J.L.F. Barbon, E. Rabinovici, ``Long time scales
and eternal black holes", Fortsch.Phys. 52 (2004) 642-649,
[arXiv:hep-th/0403268];
}

\lref\witt{ E.~Witten, ``Some comments on string dynamics,''
arXiv:hep-th/9507121.
}

\lref\vafao{ H.~Ooguri and C.~Vafa, ``Two-Dimensional Black Hole
and Singularities of CY Manifolds,'' Nucl.\ Phys.\ B {\bf 463}, 55
(1996) [arXiv:hep-th/9511164].
}

\lref\kuta{ A.~Giveon and D.~Kutasov, ``Little string theory in a
double scaling limit,'' JHEP {\bf 9910}, 034 (1999)
[arXiv:hep-th/9909110].
}

\lref\AlvarezGaumeZI{ L.~Alvarez-Gaume, J.~L.~F.~Barbon and
C.~Crnkovic, ``A Proposal for strings at D > 1,'' Nucl.\ Phys.\ B
{\bf 394}, 383 (1993) [arXiv:hep-th/9208026].
}

\lref\KorchemskyTT{ G.~P.~Korchemsky, ``Matrix model perturbed by
higher order curvature terms,'' Mod.\ Phys.\ Lett.\ A {\bf 7},
3081 (1992) [arXiv:hep-th/9205014];
G.~P.~Korchemsky, ``Loops in the curvature matrix model,'' Phys.\
Lett.\ B {\bf 296}, 323 (1992) [arXiv:hep-th/9206088].
}

\lref\Hawkingpage{ S.~W.~Hawking and D.~N.~Page, ``Thermodynamics
Of Black Holes In Anti-De Sitter Space,'' Commun.\ Math.\ Phys.\
{\bf 87}, 577 (1983).
}

\lref\horoP{ G.~T.~Horowitz and J.~Polchinski, ``A correspondence
principle for black holes and strings,'' Phys.\ Rev.\ D {\bf 55},
6189 (1997) [arXiv:hep-th/9612146].
}

\lref\barbon{ J.~L.~F.~Barbon and E.~Rabinovici, ``Closed-string
tachyons and the Hagedorn transition in AdS space,'' JHEP {\bf
0203}, 057 (2002) [arXiv:hep-th/0112173];
J.~L.~F.~Barbon and E.~Rabinovici, ``Touring the Hagedorn Ridge,''
arXiv:hep-th/0407236.
}

\lref\affleck{ I.~Affleck, ``Quantum Statistical Metastability,''
Phys.\ Rev.\ Lett.\ {\bf 46}, 388 (1981).
}

\lref\Tan{N.~Deo, S.~Jain and C.~I.~Tan, ``String Statistical
Mechanics Above Hagedorn Energy Density,'' Phys.\ Rev.\ D {\bf
40}, 2626 (1989); \quad N.~Deo, S.~Jain and C.~I.~Tan, ``The ideal
gas of strings,'' HUTP-90-A079; \quad M.~J.~Bowick and
S.~B.~Giddings, ``High Temperature Strings,'' Nucl.\ Phys.\ B {\bf
325}, 631 (1989).
}

\lref\SpradlinPP{ M.~Spradlin and A.~Volovich, ``A pendant for
Polya: The one-loop partition function of N = 4 SYM on R x S(3),''
hep-th/0408178.
}

\lref\GomezReinoPW{ M.~Gomez-Reino, S.~Naculich and H.~Schnitzer,
``Thermodynamics of the localized D2-D6 system,''
arXiv:hep-th/0412015.
}

\lref\HadizadehBF{ S.~Hadizadeh, B.~Ramadanovic, G.~W.~Semenoff
and D.~Young, ``Free energy and phase transition of the matrix
model on a plane-wave,'' arXiv:hep-th/0409318.
}

\lref\ambj{ J.~Ambjorn, B.~Durhuus and T.~Jonsson, ``Quantum
geometry. A statistical field theory approach,'' Cambridge
Monogr.\ Math.\ Phys.\ {\bf 1}, 1 (1997).
}

\lref\LiuVY{ H.~Liu, ``Fine structure of Hagedorn transitions,''
arXiv:hep-th/0408001.
}

\lref\perry{ M.~Perry, ``Instabilities in gravity and
supergravity,'' in {\it Superspace and Supergravity}, S.~Hawking
and M.~Rocek (eds). Cambridge, Cambridge University Press 1981.}

\lref\GrossPY{ D.~J.~Gross, M.~J.~Perry and L.~G.~Yaffe,
Phys.\ Rev.\ D {\bf 25}, 330 (1982).
}

\lref\vijay{ V.~Balasubramanian, K.~Larjo and J.~Simon, ``Much ado
about nothing,'' hep-th/0502111.
}

\lref\eva{ A.~Adams, X.~Liu, J.~McGreevy, A.~Saltman and
E.~Silverstein, ``Things fall apart: Topology change from winding
tachyons,'' hep-th/0502021.
}



\Title{\vbox{\baselineskip12pt \hbox{hep-th/0502227}
\hbox{CERN-PH-TH/04-251} \hbox{IFT/05/11} \hbox{MIT-CTP-3591}
\hbox{TIFR/TH/05-03}
}}%
{\vbox{\centerline{Finite temperature effective action, AdS$_5$
black holes, }
\medskip
\centerline{ and $1/N$ expansion}} }

\smallskip
\centerline{Luis Alvarez-Gaume$^\flat$, Cesar Gomez$^\natural$,
Hong Liu$^\ast$, and Spenta~R.~Wadia$^{\flat, \star}$ }
\medskip

\bigskip
\centerline{$^\flat${\it Physics Dept.-Theory Division, CERN,
CH-1211,Geneva 23, Switzerland}}
\smallskip
\centerline{$^\natural${\it Instituto de Fsica Terica,
Universidad Autnoma de Madrid, Cantoblanco 28049 Madrid, Spain}}
\smallskip

\centerline{$^\ast$ {\it Center for Theoretical Physics,
Massachusetts Institute of Technology, Cambridge, MA 02139}}
\smallskip

\centerline{$^\star${\it Dept. of Theoretical Physics, Tata
Institute of Fundamental Research, Mumbai, 400 005,
India\foot{Permanent address}}}
\bigskip

\noindent

We propose a phenomenological matrix model to study string theory
in $AdS_5 \times S_5$ in the canonical ensemble. The model
reproduces all the known qualitative features of the theory. In
particular, it gives a simple effective potential description of
Euclidean black hole nucleation and the tunnelling between thermal
AdS and the big black hole. It also has some interesting
predictions. We find that there exists a critical temperature at
which the Euclidean small black hole undergoes a Gross-Witten
phase transition. We identify the phase transition with the
Horowitz-Polchinski point where the black hole horizon size
becomes comparable to the string scale. The appearance of the
Hagedorn divergence of thermal AdS is due to the merger of saddle
points corresponding to the Euclidean small black hole and thermal
AdS. The merger can be described in terms of a cusp ($A_3$)
catastrophe and divergences at the perturbative string level are
smoothed out at finite string coupling using standard techniques
of catastrophe theory.

\bigskip

\Date{Feb 25, 2005}






\newsec{Introduction}

The AdS/CFT correspondence has enabled us to begin understanding
various aspects of quantum gravity in a more quantitative way
\refs{\malmag}. In particular reliable computations of black hole
thermodynamics have been possible using the $AdS_3/CFT$
correspondence~\refs{\malmag,\davidrep}. In this paper we would
like to address some aspects of $AdS_5$ black hole physics in the
context of type IIB string theory in $AdS_5 \times S^5 $, using
the dual gauge theory at finite temperature.

The thermodynamic aspects of quantum gravity in AdS spacetime were
discussed long ago in an important paper by Hawking and
Page~\refs{\Hawkingpage}, who realized that it is possible to
define a canonical ensemble for quantum gravity and in particular
for Schwarzschild black holes. They found that an asymptotic AdS
spacetime allows two Schwarzschild black hole solutions, which are
since called small black hole (SBH) and big black hole (BBH). As
the names suggest SBH can have a horizon radius that can be very
small compared to the size of AdS and BBH has a horizon radius
which is comparable to (or much larger than) the radius of
$AdS_5$. Further SBH has negative specific heat and is unstable,
while BBH has positive specific heat and is stable (meta-stable.)
Hawking and Page also found that the system undergoes a first
order phase transition at a temperature $T_1$ comparable to the
inverse curvature radius of the spacetime. Below $T_1$, the system
is described by a thermal gas in AdS, while above $T_1$ it is
described by a BBH. With the discovery of the AdS/CFT
correspondence~\refs{\malda,\GubserBC,\WittenQJ},
Witten~\refs{\WittenQJ,\witten} realized that a BBH in $AdS_5$ is
naturally described by the deconfinement phase of $\NN=4$
Super-Yang-Mills (SYM) theory on $S^3\times S^1$. He argued that
the Hawking-Page transition corresponds to a large $N$
deconfinement transition in the gauge theory at strong coupling.

Several authors \refs{\sundborg, \polyakov, \aharony} have
discussed the partition function of the free $\NN=4$ SYM
theory\foot{
See~\refs{\HallinKM,\FuruuchiSY,\SchnitzerQT,\FidkowskiFC,\AharonyIG,\GomezReinoPW,\SpradlinPP,
\HadizadehBF} for other recent discussions of phase transitions in
weakly coupled Yang-Mills theory.} and found that the large $N$
deconfinement transition persists at zero coupling. In particular
it was found that the deconfinement transition happens exactly at
the Hagedorn temperature of the low temperature thermal AdS
phase\foot{At strong coupling, the Hagedorn temperature for the
thermal AdS is much higher than the Hawking-Page temperature.}.
Near the Hagedorn temperature, the free energies of both high and
low temperature phases become divergent and string perturbation
theory\foot{Since the $1/N$ expansion in the boundary theory
corresponds to the perturbative string expansion in the bulk, in
this paper we will use the word ``large $N$ expansion'' and
``perturbative string expansion'' interchangeably.} breaks down.
In \refs{\LiuVY} the smoothening of the Hagedorn transition at
finite string coupling was discussed. This requires a careful
understanding of nonperturbative effects in $1/N$. It was found
that the divergences in perturbation theory are removed by two
distinct mechanisms. The divergent terms in the high temperature
phase can be resummed, leading to a noncritical string
description. This happens at a scale $T-T_H \sim N^{-{4 \ov 3}}$.
The Hagedorn divergence of the low temperature phase is removed by
summing over the contributions from the thermal AdS and the
noncritical string background, happening at $T-T_H \sim N^{-2}$.

In this paper we extend the analysis of~\refs{\LiuVY} to finite `t
Hooft coupling and study various non-perturbative aspects of black
hole physics in $AdS_5$ using the boundary gauge theory. The
latter is precisely formulated but technically difficult to deal
with in the strong coupling region, where one can make contact
with gravity. In such circumstances one is, naturally led to an
effective action approach, relying on the choice of an order
parameter and its symmetries. The difficulty is then transferred
to the coupling and temperature dependence of coefficients of the
effective action coding the microscopic theory. In spite of this
difficulty one can hope to make progress. Above all, one is
encouraged by the success of a similar programme in QCD.

The strategy has two parts. First one may try to extract certain
{\it universal} features of string theory in $AdS_5$ from the
effective action. The hope is that universal features do not
depend on the exact details of the effective action and can be
extracted by exact analysis of a tractable model. Secondly one can
approximately determine the coefficients of the effective action
by explicitly matching, in our case, with data in the dual
supergravity description.

String theory backgrounds like thermal AdS, BBH and SBH appear as
saddle points in the Euclidean path integral of Yang-Mills theory.
Perturbative string expansion around each of them is given by the
large $N$ expansion around the corresponding saddle point in the
boundary theory\foot{For the perturbative $1/N$ expansion around
the saddle point, it is not important whether the saddle of
interest is stable, metastable or unstable.}. As one varies the
temperature, such expansions break down at various places where
their coefficients develop nonanalytic behavior. One example is
the Hagedorn temperature of thermal AdS. The other is the
temperature (called $T_0$ by~\refs{\Hawkingpage}) at which SBH and
BBH saddles appear. While such nonanalytic behavior is rather
puzzling and hard to understand from the perturbative string point
of view, in the dual Yang-Mills theory they arise due to the fact
that in the large $N$ limit the number of degrees of freedom goes
to infinity. The non-analyticity occurs for the same reason as in
the thermodynamic limit of classical statistical physics. At these
temperatures, as in the case analyzed in~\refs{\LiuVY}, a
non-perturbative treatment is required {\it no matter how large
$N$ is}. Thus these non-analyticities are excellent probes of the
non-perturbative structure of the theory. Moreover, since their
appearance is intrinsically tied to the large $N$ limit, it is
expected thnat they possess a certain degree of universality, just
as in critical behavior in condensed matter systems. One may hope
that the {\it qualitative} behavior at these critical temperatures
sould be insensitive to the precise details of the theory and
could be captured by studying much simpler systems. This gives us
hope that we can study critical behavior, and hence
non-perturbative aspects of large $N$ Yang-Mills theory at finite
or strong coupling, and in turn yielding insights into string
theory in $AdS_5$.

It was discussed in~\refs{\aharony} that the partition function of
the SYM theory can be written as a matrix integral over the
effective action of the Polyakov loop. For Yang-Mills theory at
finite coupling, we have no way of computing this effective action
explicitly. Nevertheless, with universality in mind, here we
propose a class of effective actions as  ``phenomenological
models'' to approximate the full theory. We show that models in
the class have a large $N$ phase structure resembling that of a
weakly coupled string theory in $AdS_5 \times S_5$. This gives
strong indication that strongly coupled $\NN=4$ SYM theory belongs
to the same universality class. This also gives us reason to
believe that critical behaviors of the bulk string theory at
places where string perturbation theory breaks down can indeed be
captured by much simpler models.

The simplest model within the class, which we will refer to as
$(a,b)$ model, can be considered as a truncation of the full
effective action of the theory to the lowest nonlinear terms\foot{
The model contains two parameters $(a,b)$, both are functions of
the 't Hooft coupling $\lam$ and temperature $T$. We will assume
some qualitative dependence of $(a,b)$ on $T$ as part of the
phenomenological input data. This model has been briefly discussed
earlier in the mean field approximation as a toy model for weakly
coupled gauge theories in~\refs{\aharony}.}. Being exactly
solvable to all order in $N$, this model provides an ideal
representative to study the critical behaviors of the universality
class. We proceed to perform a detailed study of various
non-perturbative aspects of this model. The results, when
translated into the language of bulk string theory, can be
summarized as follows:

\item{1.} We give an effective potential description of the
tunnelling between thermal AdS and the BBH. The Euclidean SBH
plays the role of the bounce (also called a thermalon). We compute
the tunnelling rate in our effective theory.

\item{2.} We find that the Euclidean SBH undergoes a third order
Gross-Witten~\refs{\gw,\wadiapre,\wadia} phase transition in the
large $N$ limit at a temperature $T_c$ below the Hagedorn
temperature. We identify the phase transition with the
Horowitz-Polchinski correspondence point~\refs{\horoP} where the
event horizon size of the SBH becomes comparable to string scale.

\item{3.} The breakdown of the perturbative string expansion of
thermal AdS at the Hagedorn temperature is due to the merger of
the saddle points corresponding to the SBH and thermal AdS. The
merger can be described in terms of a cusp ($A_3$) catastrophe.
The simplest possibility allowed by the symmetry. Similarly the
breakdown of perturbative string theory around the BBH when it
merges with the SBH saddle can be understood in terms of a fold
($A_2$) catastrophe. The divergences at the perturbative level are
smoothed out at finite $N$ using the standard techniques of
catastrophe theory.

\item{4.}A common theme in our study of the critical behavior when
a CFT approaches a singular point is that there always exists a
double scaling limit and it is likely that the theory in the
double scaling limit is described by a noncritical string
background. This also resonates with the result of~\refs{\LiuVY}
and the behavior of other singular CFTs discussed
in~\refs{\vafao,\witt,\kuta}.

\ndt While these features are studied explicitly only in the
simplest $(a,b)$ model, we believe they persist for all models in
the class due to universality of the large $N$ phase transition
and the catastrophe.

The plan of the paper is as follows. In the next section we review
some aspects of the thermodynamics of quantum gravity in $AdS_5
\times S_5$ which we aim to reproduce in the large $N$ limit of
our ``phenomenological'' models. In section 3, we review some
aspects of the computation of the Yang-Mills partition function
using the effective action of the Polyakov loop and present the
truncated models. Section 4 is devoted to a detailed study of the
phase structure of the $(a,b)$ model  at large $N$. We discuss in
detail the thermal history of the theory in the canonical
ensemble. We also show that the sharp Hawking-Page transition in
supergravity is smoothed out to a finite cross region at finite
$N$. In section 5 we elucidate the role of SBH as the bounce which
mediates the tunnelling between BBH and thermal AdS (and vice
versa depending on the temperature) and calculate the tunnelling
rate. We also connect the bounce and the large order behavior of
perturbative theory. In section 6 we study the critical behaviors
of the theory at temperatures where the perturbative string
expansion breaks down around at least one of the three
backgrounds. They can be understood using catastrophe theory. We
conclude in section 7 with a discussion of future directions. We
have also include a few appendices which contain details of some
calculations.


\newsec{Review of Hawking-Page transition in Euclidean Quantum
gravity}

In this section we review the results of~\refs{\Hawkingpage}, to
be reproduced using the matrix models in later sections.

The canonical ensemble for quantum gravity in AdS can be defined
as a path integral over the metric and all other fields asymptotic
to AdS with time direction periodically identified with a period
$\beta = 1/T$. At semi-classical level, i.e. $R^2/l_p^2 \gg 1$,
where $R$ is the curvature radius of AdS, such a path integral is
dominated by configurations near the saddle points, i.e. classical
solutions to the Einstein equations. If we assume spherical
symmetry and zero charge, there are three possible critical
points, which are thermal AdS$_5$ (Euclidean AdS with time
direction periodically identified), a big (Schwarzschild) black
hole (BBH) and a small black hole (SBH). Among them thermal AdS
and BBH are locally stable, while SBH has a negative mode and it
is unstable. The thermal AdS background has topology $S^1 \times
R^4$, while SBH and BBH have topology $R^2 \times S^3$, all of
them with a common boundary $S^1 \times S^3$. The Euclidean time
direction in black hole backgrounds are contractible and the
winding numbers are not conserved. In contrast the time circle in
thermal AdS is noncontractible and the winding number is
conserved.

The classical action for thermal AdS is $I_1 = 0$. This is
standard in string theory: with a noncontractible time circle,
there is no genus zero contribution to the free energy. A
Schwarzschild black hole solution exists in AdS only for a Hawking
temperature greater than \eqn\criTs{ T_0 = {\sqrt{2} \ov \pi R },
\qquad \beta_0 = {1 \ov T_0} = {\pi R \ov \sqrt{2}} } For $T >
T_0$, there are two possible black holes, whose horizon sizes are
given by \eqn\temper{ {r_{+} \ov R} = {1 \ov \sqrt{2}}
\left[{\beta_0 \ov \beta} \pm \sqrt{{\beta_0^2 \ov \beta^2}-1}
\right] } The corresponding classical Euclidean action is given by
\eqn\exacR{ I = {R^3 \ov 2 \kappa^2} 2 \pi \Om_{3} \le({r_+ \ov
R}\ri)^{3} {1 - \le({r_+ \ov R}\ri)^2 \ov 1 + 2 \le({r_+ \ov
R}\ri)^2 } \ , } where $2 \kappa^2$ is the five-dimensional
Newton's constant\foot{Note that ${R^3 \ov 2 \kappa^2} \propto
N^2$, where $N$ is the rank of the gauge group in the boundary
Yang-Mills theory.}. We will denote $I_+, I_-$ the classical
actions for large and small black hole respectively. The specific
heat of the large black hole is positive and thus it is
thermodynamically stable (i.e. it can reach locally stable thermal
equilibrium with thermal radiation). The small black hole has a
negative specific heat. The action $I_-$ of the small black hole
is always greater than the action of thermal AdS and of the big
black hole. At temperature \eqn\HawkpT{ T_1 = {3 \ov 2 \pi R} >
T_0 } the action for the big black hole is $I_+ =0 = I_1$. When
$T_0 < T < T_1$, $I_+ > 0$, and the saddle corresponding to
thermal AdS dominates. When $T > T_1$, $I_+ < 0$, the big black
hole (BBH) dominates. There is a change of dominance at $T_1$.
This is the Hawking-Page transition. In the classical limit
$\kappa^2 \to 0$, this is a sharp first order transition. We
expect that at finite $\kappa^2$ the transition should be smoothed
out. This we will see explicitly in the gauge theory description.

In the Minkowski description the spectrum of fluctuations around
$AdS_5$ and BBH (all of positive frequency in Euclidean and
Minkowski descriptions) can be interpreted in terms of physical
particles and they constitute the meta-stable thermal ensemble
around these backgrounds. The SBH on the other hand has a negative
eigenvalue in the spectrum of small fluctuations in the Euclidean
description. Hence the SBH fits the description of an instanton
relevant for the tunnelling between thermal $AdS_5$ and the BBH.
For example, one expects the rate for a BBH to tunnel into a
thermal AdS is expected to be \eqn\probI{ \Ga_1 = A_1 e^{- (I_- -
I_+)} } That is, through thermal fluctuation, a big black hole can
turn into a small black hole. The small black hole (since it has a
negative specific heat) then can either shrink to thermal AdS by
emitting thermal radiation or grow back into a big black hole by
absorbing radiation. Similarly, the thermal AdS background has
also a nonzero probability to nucleate a small black hole with
probability \eqn\probII{ \Ga_2 = A_2 e^{- I_- } } Afterwards the
small black hole can shrink back to the thermal AdS or grow into a
big black hole. The prefactors $A_1, A_2$ in \probI\ and \probII\
are given by the determinants of small fluctuations around the
relevant background. In thermal equilibrium, the probability to go
from a typical state in thermal AdS to that of a big black hole or
back should be the same.

When $T_0 < T < T_1$, the big black hole phase is metastable,
since it has a higher free energy than that of that of thermal AdS
and $\Ga_1 > \Ga_2$. But string perturbation theory around it is
well defined until $T_0$ is reached where we expect the
perturbation theory to break down. Similarly, when $T > T_1$,
thermal AdS becomes metastable and $\Ga_2 > \Ga_1$. For a large
AdS with $R \gg l_s$ ($l_s$ is the string length) the perturbation
theory around thermal AdS breaks down at a much higher Hagedorn
temperature $T_H \sim {1 \ov l_s}$. In the Hawking-Page
discussion, there also exists a temperature $T_2$ beyond which the
thermal graviton gas in AdS will collapse into a big black hole.
For a weakly coupled string theory in $AdS_5 \times S_5$, $T_2$ is
of order ${1 \ov (R l_p^4)^{1 \ov 5}}$ and is much higher than the
Hagedorn temperature $T_H \sim {1 \ov l_s}$ for thermal AdS.

\newsec{Effective action at finite temperature}

In this section we will introduce a phenomenological matrix model
for understanding string theory in $AdS_5 \times S_5$ at finite
temperature.

We first give some general discussion of the partition function of
$\NN=4$ SYM theory on $S^3$. We consider the theory in the
canonical ensemble, i.e. the Euclidean time direction is
periodically identified with a period of $\beta = {1 \ov T}$. It
was pointed out in~\refs{\aharony} (see also~\refs{\pisarski})
that the Yang-Mills theory partition function on $S^3$ at a
temperature $T$ can be reduced to an integral over a unitary
$U(N)$ matrix $U$, which is the zero mode of Polyakov loop on
$S^3$,
 \eqn\utri{
  \ZZ (\lam,T)= \int dU \, e^{ S_{eff} (U)}
   }
with
 \eqn\uniWil{
 U = P \exp \left(i \int_0^{ \beta} A d \tau \ri)
 }
 where $A (\tau) $ is the zero mode of the time component of the
gauge field in $S^3$. This follows from the fact that apart from
$A$ all modes of $\NN=4$ SYM on $S^3$ are massive. Hence one can
integrate them out to obtain an effective action for $A$. Gauge
invariance requires that the effective action must be expressed in
terms of products of $\tr U^n$ with $n$ an integer, since these
are the only gauge invariant quantity that can be constructed from
$A$ alone. $S_{eff} (U)$ has a $Z_N$ symmetry
$$
U \to e^{{2 \pi i \ov N}} U \
$$
due to global gauge transformations which are periodic in the
Euclidean time direction up to $Z_N$ factors. A generic term in
$S_{eff} (U)$ will have the form
$$
\tr U^{n_1} \tr U^{n_2} \cdots \tr U^{n_k}, \qquad n_1 + \cdots
n_k = 0 \; ({\rm mod} \; N), \qquad k>1
$$
We can expand $S_{eff}$ in terms of a complete set of such
operators, with the first few terms
 \eqn\unexp{
  S_{eff} (U) = a_1
\tr U \tr U^{-1} + b_1 \le(\tr U \tr U^\dagger\ri)^2 + a_2 \tr U^2
\tr U^{-2} + c_1 \tr U^2 \tr U^\dagger \tr U^\dagger + \cdots
 }
The coefficients in the expansion are functions of 't Hooft's
coupling $\lam$, and $T$. While these coefficients are in
principle calculable at weak coupling, the explicit computations
are in general very complicated (see e.g.~\refs{\minW}). At finite
or large `t Hooft coupling, there is no available tool at the
moment to attempt such a computation. Even one were able to find
the expansion \unexp\ explicitly, to perform a finite $N$
computation of the matrix integral \utri\ is still a daunting
task, if not impossible.

In order to make progress, in this paper, we will consider the
truncation of \unexp\ to terms containing only powers of $\tr U
\tr U^{-1}$, i.e. we consider an effective action of the form
 \eqn\adjpoly{\eqalign{
 S_{eff} (U)
 & = S(x), \qquad x = {1 \ov N^2 } \tr U \tr U^\dagger \cr
 & = a \tr U \tr U^{-1} + {b \ov
 N^2} \le(\tr U \tr U^\dagger\ri)^2 + {c \ov N^4} \le(\tr U \tr
 U^\dagger\ri)^3 +
 \cdots \cr
 }}
Our consideration is  phenomenological, motivated by the AdS/CFT
correspondence to search for effective actions which lie within
the same universality class as that of the SYM theory at finite
coupling.  At a heuristic level, one may also consider \adjpoly\
as arising from \unexp\ by ``integrating out'' all the higher
moments $\tr U^{n}$, $\tr U^{-n}$ ($n >1$). In~\refs{\pisarski} it
was also argued based on group theory considerations that terms of
the form \adjpoly\ dominate over other types terms in \unexp\ in
the large $N$ limit.

We will consider a class of matrix models of the form \adjpoly\
satisfying the conditions that $S(x)$ is convex and $S'(x)$ is
concave. We show in the next section and in Appendix A that in this
class the large $N$ phase structure appears to be universal. In
particular, the phase structure precisely reproduces\foot{We will
assume some qualitative dependence of $S(x)$ on $T$ as part of the
phenomenological input data.} the phase structure of a weakly coupled string theory in AdS. We believe this is a strong indication that
strongly coupled $\NN=4$ SYM theory also lies in this class.
The simplest model in this universality class is given by the
first two terms in~\adjpoly
 \eqn\themodel{
 \ZZ(a,b) = \int d U \, \exp\le[a \tr U \tr
 U^\dagger + {b \ov N^2} \le(\tr U \tr U^\dagger \ri)^2 \ri]
  }
with $b >0$.

As discussed in the introduction, we are interested in the
critical behaviors in regions of parameter space where large $N$
expansions around various saddle points break down. Being exactly
solvable at finite $N$, \themodel\ provides a simple, nice
representative to study the critical behaviors of the universality
class.

To conclude this section, we note that \themodel\ was discussed
in~\refs{\aharony} in the Hartree-Fock approximation. It was noted
that for $b> 0$ a first-order phase transition resembling the
Hawking-Page transition occurs. This observation was a motivation
for the investigations in this paper.


\newsec{Large $N$ phase structure of the universality class}

In this section we study the large $N$ phase structure of
\adjpoly\ and \themodel\ ($b>0$). For our later purpose of
studying the critical behaviors of \themodel, we give a detailed
discussion of the phase structure of \themodel\ using a method
suitable for finite $N$ analysis. We point out general  matrix
model \adjpoly
 \eqn\uhgm{
 \ZZ = \int dU \, e^{N^2 S(x)} ,
 \qquad x = {1 \ov N^2} \tr U \tr U^\dagger \
  }
has the same large $N$ phase structure as \themodel\ provided that
$S(x)$ is convex and $S'(x)$ is concave. For completeness we have
included in Appendix A an alternative discussion of the phase
structure of \uhgm\ using the Hartree-Fock method.

\subsec{Effective potential}

{}For $b>0$, equation \themodel\ can be rewritten using a Lagrange
multiplier $\mu$: \eqn\furmo{\eqalign{ \ZZ (a,b) & = {N \ov 2
\sqrt{\pi b}} \int_{-\infty}^\infty d \mu \, e^{-{N^2 \ov 4b}
(\mu-a)^2} \int dU \, \exp \le[\mu \tr U \tr U^\dagger\ri] \ . \cr
}} The matrix integral in \furmo\ can be further simplified by
introducing another Lagrange multiplier $g$. For example for $\mu
0$, one finds \eqn\comPa{\eqalign{ e^{N^2 \FF (\mu)} & = \int dU
\, \exp \le[\mu \tr U \tr U^\dagger\ri] \cr & = {N^2 \ov 2 \mu}
\int_0^\infty dg \, g \, e^{-{N^2 g^2 \ov 4 \mu} + N^2 F (g)} \cr
}} with \eqn\Hjho{ e^{N^2 F(g)} = \int d U \, \exp \le[{N g \ov 2}
\le(\tr U + \tr U^\dagger \ri) \ri] } The formula for $\mu < 0$ is
obtained by taking $g \to i g$.

The large $N$ expansion of \Hjho\ and the corresponding third
order phase transition is well known ~\refs{\gw,\wadiapre,\goldsm}
\eqn\gtfp{ F (g) = \cases{{g^2 \ov 4} + {\rm nonperturbative} & $g
\leq 1 \quad {\rm or} \quad g \; {\rm imaginary}$ \cr\cr g - \ha
\log g - {3 \ov 4} + O(1/N^2) & $g > 1 $ \cr }} The order
parameter of \Hjho\ can be taken to be \eqn\tyiD{\eqalign{ \rho_1
(g) & = {1 \ov N} \vev{\Tr U}_g = {1 \ov N} \vev{\Tr U^\dagger}_g
= {\p F \ov \p g} \cr & = \cases{{g \ov 2} + \cdots & $ g < 1
\quad {\rm or} \quad g \; {\rm imaginary} $ \cr\cr 1-{1 \ov 2g} +
\cdots & $ g > 1$ \cr } \cr } } characterizing the eigenvalue
distribution of $U$. When $0 \leq \rho_1 < \ha$ ($g <1$), the
system is in a phase whose eigenvalue distribution does not have a
gap on the unit circle. In particular, for $g=0$, the distribution
is uniform. When for $1 > \rho_1 > \ha$ ($g >1$), the distribution
develops a gap. \gtfp\ and \tyiD\ do not apply to $g \approx 1$,
where the system undergoes a third order phase
transition~\refs{\gw} in the large $N$ limit. At finite $N$ the
third order discontinuity in \gtfp\ is smoothened out by
non-perturbative effects. They will be discussed in later sections
when needed.

One can now rewrite \themodel\ as a two dimensional integral
\eqn\yhios{ \ZZ = {N^3 \ov 4 \sqrt{\pi b}} \int_{-\infty}^\infty
{d \mu \ov \mu} \int_0^\infty g dg \, e^{-N^2 V(\mu, g)} } with
\eqn\defVg{ V (\mu,g) = \cases{ {1 \ov 4b} (\mu-a)^2 - {g^2 \ov
4}{1-\mu \ov \mu} & $ \mu < 0$ \cr\cr {1 \ov 4b} (\mu-a)^2 + {g^2
\ov 4}{1-\mu \ov \mu} & $ \mu > 0,\; 0 \leq g < 1$ \cr\cr {1 \ov
4b} (\mu-a)^2 + {g^2 \ov 4 \mu} - g + \ha \log g + {3 \ov 4} +
O(1/N^2) & $\mu > 0, \; g > 1$ \cr }} It is often convenient to
integrate out $g$ to reduce \yhios\ to a one dimensional integral
\eqn\furmoa{ \ZZ(a,b) = {N \ov 2 \sqrt{\pi b}}
\int_{-\infty}^\infty d \mu \, e^{-N^2 Q (\mu)} } with \eqn\DFQ{ Q
(\mu) = {1 \ov 4b} (\mu-a)^2 - \FF(\mu) } and $\FF (\mu)$ was
defined in \comPa. The large $N$ expansion for $\FF(\mu)$ was
found in~\refs{\LiuVY}, e.g. the leading order terms are
\eqn\resff{ \FF (\mu) = \cases{ 0 - {1 \ov N^2} \log (1-\mu) +
\cdots & $\mu < 1$ \cr\cr {1 \ov 2} {w \ov 1 - w} + \ha \log (1 -
w) + O\le({1 \ov N^2} \ri) & $\mu > 1$ \cr }} where for $\mu >1$
we have introduced \eqn\DEFw{ w =\sqrt{1-{1 \ov \mu}} \ . }

Inherited from \gtfp, \defVg\ has a third order discontinuity at
$g=1$ which needs to be supplemented with a non-perturbative
treatment. \DFQ\ with $\FF$ given by \resff\ has
divergences\foot{The second and higher derivatives of $\FF (\mu)$
are also divergent for $\mu \to 1_+$.} and first order
discontinuity at $\mu =1$. Again a non-perturbative treatment is
necessary, as discussed in detail in~\refs{\LiuVY}. Part of the
subtlety at $\mu=1$ in \DFQ\ has to do with that at $\mu =1$, $g$
becomes massless and the effective potential $V(\mu,g)$ is flat in
the range $0<g<1$. Thus near $\mu =1$ it is more convenient to use
the two dimensional effective potential \defVg. Also in \resff\ we
have suppressed a subdominant term which should be taken into
account in the analysis of the phase structure. This is
automatically taken care of in the two-dimensional integral
\yhios. In this and following sections we will use both forms of
the effective potential \defVg\ and \DFQ\ depending on
convenience. The one-dimensional effective potential $Q(\mu)$ is
easier to visualize than the two-dimensional potential $V
(\mu,g)$. But as we mentioned, equation \defVg\ is more convenient
around $\mu=1$.

We note that the trick used in \furmo\ to reduce \themodel\ to an
integral transform of \comPa\ can be generalized to find the large
$N$ phase structure of general matrix models \uhgm\ with $S(x)$
convex.  Since $S(x)$ is convex, it admits a Legendre
transform\foot{ For general properties of Legendre transformations
see for instance the book \refs{\arnold}.}:
$$
\SS(\mu)=Max_x(\mu x-S(x))
$$
The Legendre transform is involutive. If we do it twice we get
back the same function. Hence $\SS(\mu)$ is also convex. We can
then write \uhgm\ as \eqn\morgh{ \ZZ= \int dU \; e^{N^2 S(x)}=\int
dU \int d\mu \; e^{N^2(\mu x -\SS(\mu))} } and the second integral
over $\mu$ is carried out using saddle points. For large $N$ this
will give an excellent approximation. Convexity in fact guarantees
that there is a unique saddle point contributing. If we now
exchange the order of integration in \morgh\ and use \comPa, we
find that \eqn\ghsn{ \ZZ = \int d \mu \, e^{- N^2 \QQ(\mu)} }
where \eqn\Deghk{ \QQ(\mu) = \SS (\mu) - \FF (\mu) } and
$\FF(\mu)$ is given by \resff\ in the large $N$ limit. We will
show in next subsection that $\QQ (\mu)$ leads to the same large
$N$ phase structure as \DFQ\ provided $\SS'(\mu)$ is also
convex\foot{That $\SS'(\mu)$ is convex means that $S'(x)$ is
concave.}.

\subsec{Phase structure}

In the large $N$ limit the critical points of $V$ in \yhios\
describe different phases of the theory which in turn correspond
to different bulk string theory geometries. The minima correspond
to (meta)stable phases, while saddle points\foot{By saddle here we
refer to saddle points of $V(\mu,g)$ on the real $\mu-g$ plane.}
(or maxima) to unstable phases. Note that in the large $N$ limit,
the eigenvalue distribution of the Polyakov loop $U$ at a critical
point follows from \Hjho\ with $g$ given by its value at the
critical point. Since there is a one-to-one map \tyiD\ between
$\rho_1$ and $g$, $g$ can be considered as the order parameter of
the theory. After integrating out $g$ one can also interpret $\mu$
in \furmoa\ as the order parameter (at least in the range $\mu
1$).

Before discussing the critical points of the theory in detail
(which is somewhat involved), we note that qualitative features of
the critical point structure of \themodel\ can be conveniently
visualized by plotting the one dimensional effective potential
$Q(\mu)$ \DFQ\ in the large $N$ limit. Depending on the values of
$(a,b)$, $Q$ can have one or three critical points (see fig. 2).
The critical point structure in the $(a,b)$ plane is plotted in
fig.~1 (see below). Below curve I in fig.~1, $Q$ has one minimum.
Between curve I and curve $H$, it has three critical points $\mu_1
< \mu_2 < \mu_3$, with two minima ($\mu_1$ and $\mu_3$) and one
maximum ($\mu_2$). The two minima change dominance on curve II. On
curve I, $\mu_2$ and $\mu_3$ merge. At curve $H$, $\mu_1$ and
$\mu_2$ merge. To the right of curve $H$, in addition to $\mu_3$,
\defVg\ also has a tachyonic saddle which is not visible in the
leading order $Q$-plot here.

\ifig\contour{This figure plots the critical point structure of
the theory in the $a-b$ plane. Below line I, there is one critical
point. There are three critical points between line I and line
$H$, two minima, one maximum. At line II, two minima exchange
dominance.} {\epsfxsize=6cm \epsfbox{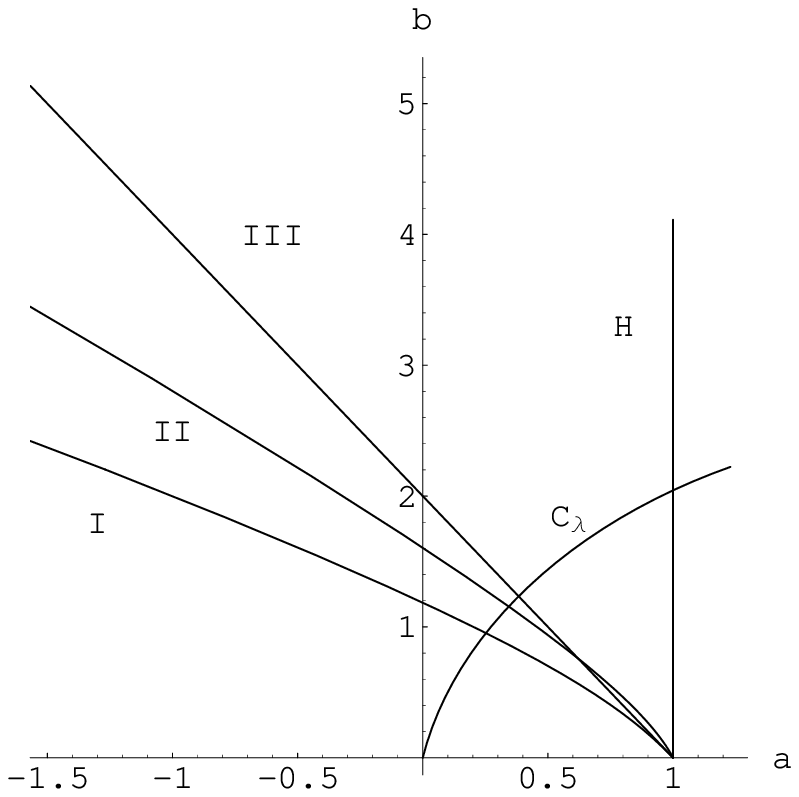}}

\ifig\Qshape{Plot 2a corresponds to $(a,b)$ below curve I of
fig.~1, 2b to $(a,b)$ between curve I and curve II, 2c to $(a,b)$
between curve II and curve $H$, and 2d to $(a,b)$ lying to the
right of curve $H$. Note the discontinuity in the first derivative
at $\mu=1$ is due to the large $N$ approximation. It is smoothened
out by non-perturbative effects.} {\epsfxsize=10cm
\epsfbox{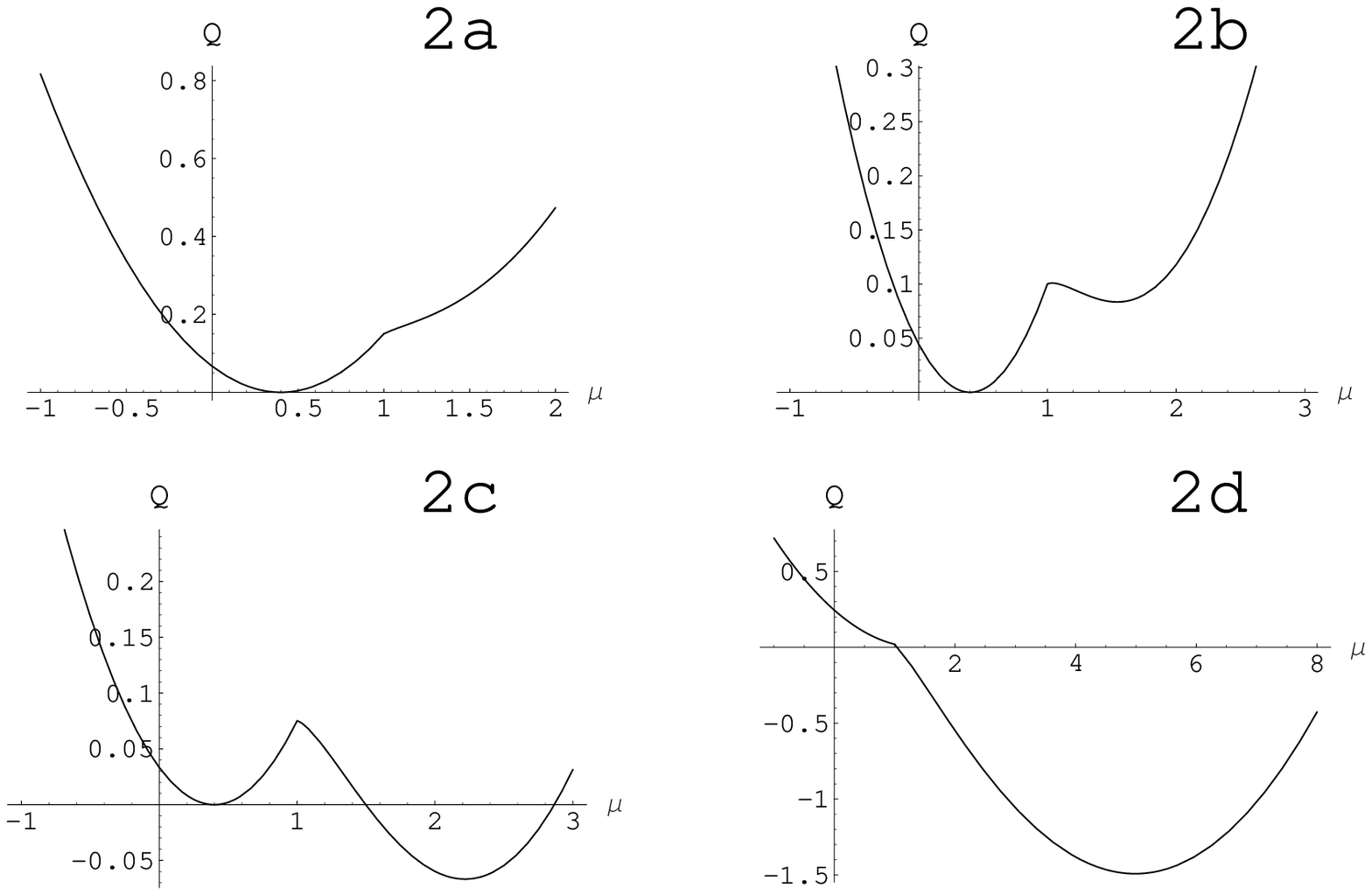}}

We now describe the critical points of \defVg\ in detail:

\item{1.} From the first two lines of \defVg\ one finds the
following critical point \eqn\theads{ \mu_1 = a , \qquad g_1 = 0 }
with \eqn\byip{ V(\mu_1, g_1) = 0 } and \eqn\sevg{ V'' =
\pmatrix{V_{\mu \mu} & V_{\mu g} \cr V_{\mu g} & V_{gg} \cr} = \ha
\pmatrix{{1 \ov b} & 0 \cr 0 & {1 -a \ov a} \cr} } where $V_{gg}$
denotes ${\p^2 V \ov \p g^2}$ and so on. For $a<1$, it is a local
minimum. $V''$ becomes singular for $a=1$ and tachyonic for $a >
1$. Since $g_1=0$, it describes a uniform eigenvalue distribution
in the unit circle.

\item{2.} For $a < 1$ and $ c= {2 (1-a) \ov b} < 1$, there is an
additional saddle point from the second line of equation \defVg\
at
 \eqn\theyo{
 \mu_2 = 1, \qquad g_2 = \sqrt{c}, \qquad c= {2
(1-a) \ov b} < 1
 } with
  \eqn\buiop{ V (\mu_1, g_1) = {(1-a)^2 \ov
 4b}
  } and
   \eqn\sert{
   V'' = \ha \pmatrix{{1 \ov b} + c & -\sqrt{c}
 \cr -\sqrt{c} & 0 \cr}
 } Note that $V''$ has a negative
eigenvalue. Since $\rho_1 (\mu_2,g_2) = \ha \sqrt{c} < \ha $,
$\mu_2$ describes a gapless phase in the eigenvalue distribution
of $U$. As $a \to 1$, \theads\ and \theyo\ merge, after which
\theyo\ disappears while \theads\ becomes tachyonic.

\item{3.} From the third line of \defVg\ the equations for the
critical points are given by \eqn\firDQg{ {1 \ov 2b} (\mu -a) -
{g^2 \ov 4 \mu^2} = 0, \qquad {g \ov 2 \mu} - 1 + {1 \ov 2 g} = 0
} Note that the second equation in \firDQg\ only has real
solutions for $\mu > 1$, in which case one finds that\foot{The
other real solution has $g <1$ and so is discarded.} \eqn\solui{
g = {1 \ov 1-w}, 
} where $w$ was introduced in \DEFw. The eigenvalue distribution
for such a $g$ is given by \eqn\whuo{ \rho_1 = 1- {1 \ov 2g} = {1
+ w \ov 2} > \ha \ . } Substituting \solui\ into the first
equation of \firDQg\ one finds an equation for $\mu$,
 \eqn\critone{
  {\mu-a \ov 2 b}  = {\le(w +1 \ri)^2 \ov 4}
   }
   Of course
\critone\ can also be obtained directly by extremizing the leading
order term of the effective potential \DFQ\ for $\mu > 1$, namely
\eqn\egst{ Q (\mu) = V(\mu, g(\mu)) = {(\mu-a)^2 \ov 4b} - {1 \ov
2} {w \ov 1 - w} - \ha \log (1 - w) + O\le({1 \ov N^2} \ri) } Note
that \eqn\secdQ{ Q''(\mu) = - {1 \ov 4} \le (-{2 \ov b} + {1 \ov
\mu^2} + {1 \ov \sqrt{\mu -1} \mu^{3/2}} \ri) } Deriving $\mu$
from \critone, $g$ and $\rho_1$ can then be found from \solui\ and
\whuo. Since $\rho_1$ is a monotonic function of $\mu$, one can
treat $Q(\mu)$ as an effective potential for $\rho_1$.

\item{4.} Equation \critone\ can be easily solved by consider the
intersections of two functions $f_1 (\mu) = {\mu-a \ov 2b} $ and
$f_2 (\mu) = {(1+w)^2 \ov 4}$. Note that $f_2$ is concave in the
range $\mu \in [1, \infty)$, while $f_1$ is a straight line. Thus
\critone\ can have at most two real solutions in the allowed
range. The result is as follows. Below curve I in \contour\ on the
$a-b$ plane, which is determined by \eqn\curvIa{ Q'(\mu) =0,
\qquad {\rm and} \qquad Q''(\mu) =0 \ , } \critone\ has no
solutions in the desired range. Between curve I and the straight
line (curve III in \contour)
 \eqn\curvIII{
  c = {2 (1-a) \ov b}=1 \ ,
  } there are two
solutions\foot{We do not give their explicit expressions in terms
of $(a,b)$, since they are complicated and not illuminating. We
will specify their qualitative behavior below.} $1< \mu_2 <
\mu_3$. It can be checked from \secdQ\ that $\mu_2$ is a maximum
of $Q$ with $Q''(\mu_2) < 0$ while $\mu_3$ is a local minimum with
$Q''(\mu_3) > 0$. At curve I, $\mu_2$ and $\mu_3$ merge and move
into the complex plane below curve I. As one approaches \curvIII\
from below ($c \to 1_+$), $\mu_2 \to 1_+$. Above line $c= 1$,
$\mu_2$ moves outside the $\mu > 1$ region to becomes \theyo\ and
\critone\ has only one solution $\mu_3$.

\item{5.} Below curve III, $\rho_1 (\mu_2) > \ha$. Above curve
III, $\mu_2$ becomes \theyo\ with $\rho_1 (\mu_2) < \ha$. Thus
$\mu_2$ undergoes a Gross-Witten type phase transition in the
large $N$ limit. We will show in section 5.2 that the transition
is precisely the third order Gross-Witten transition.

\item{6.} There is an additional curve (curve II in \contour), in
the $a-b$ plane, determined by equation \eqn\hawPl{ Q'(\mu_3) =0,
\qquad {\rm and} \qquad Q(\mu_3) =0, } where the two minima
$\mu_1$ and $\mu_3$ become of equal height
$$
0 = V(\mu_1 , g_1) = V(\mu_3, g_3) < V(\mu_2, g_2) \ .
$$
Below curve II, one has
$$
0 = V(\mu_1 , g_1) < V(\mu_3, g_3) < V(\mu_2, g_2) \ ,
$$
and above curve II
$$
V(\mu_3, g_3) < 0 = V(\mu_1 , g_1) < V(\mu_2, g_2) \ .
$$

\ndt To summarize, the structure of critical points for \yhios\ in
the $a-b$ plane is plotted in \contour. Below curve I \curvIa, $V$
has a unique minimum \theads. Between curve I and line $a=1$
(curve $H$ in \contour), there are three critical points $\mu_1 <
\mu_2 < \mu_3$ with $\mu_1, \mu_3$ minima, while $\mu_2$ a saddle
point in the $\mu-g$ plane with one negative eigenvalue. At curve
I, $\mu_2$ and $\mu_3$ merge together. At curve II, $\mu_1$ and
$\mu_3$ exchange dominance and the system has a first order phase
transition. At curve III, the saddle point $\mu_2$ undergoes a
Gross-Witten phase transition. At the vertical line $H$, $\mu_1$
and $\mu_2$ merge together. To the right of line $H$, $\mu_1$
becomes tachyonic, $\mu_2$ disappears, and $\mu_3>1$ remains a
minimum. Note that curve II and curve III always lie between curve
I and line $a=1$, but they can be above or below each other. Close
to $a=1,b=0$, line III lies below curve II and then intersects
with and rises above it.

\subsec{Critical points for general models}

We now consider the critical point structure of the general
effective action \uhgm\ and \ghsn. As is clear from the derivation
above, the overall phase structure of \DFQ\ presented in \Qshape\
to a large extent only depends on the convexity of the function
${(\mu-a)^2 \ov 4b}$ and its derivative. Our discussion above for
\DFQ\ goes through for a convex $\SS(\mu)$ in \Deghk\ provided
that $\SS'(\mu)$ is also convex. For example, for $\mu<1$, with
$\FF (\mu)$ given by \resff, $\QQ(\mu)$ has just one critical
point given by the minimum of $\SS(\mu)$. For $\mu > 1$, the
critical points of \Deghk\ satisfy the equation (which is a
generalization of \critone)
 \eqn\newcirr{
 \SS ' (\mu) = {(1+w)^2 \ov 4}
 }
Since the right hand side of \newcirr\ is concave and $\SS'(\mu)$
is convex, $\QQ$ can have at most two critical points in the range
$\mu \in (1, \infty)$. The pattern and the evolution of the
critical points with the parameters of the theory also precisely
resemble that of \DFQ\ including a Gross-Witten  phase transition
for $\mu_2$ at $\mu=1$. Note that convexity of $\SS'(\mu)$ implies
that $S'(x)$ is concave. We thus conclude that the structure of
critical points of \uhgm\ is {\it universal} if $S(x)$ is convex
and $S'(x)$ is concave. In appendix A, we give a more detailed
description of the phase structure of \uhgm\ using the
Hartree-Fock approximation.

\subsec{Thermal History}

As discussed in section 3, we would like to use the matrix model
\themodel\ as a phenomenological model to study weakly coupled
string theory in $AdS_5 \times S_5$ at finite temperature. The
parameters $a,b$ are functions of the 't Hooft coupling $\lam$ and
the temperature $T$. In the last subsection, we analyzed its large
$N$ critical point structure. In this subsection we will show that
with some weak assumptions about the $\lam$ and $T$ dependence of
$a,b$, the model captures all the essential features of the bulk
story, which we reviewed in sec. 2.

We first identify the critical points of $V$ in the large $N$
limit with the saddle points of the Euclidean gravity. $\mu_1 =a$
has $\rho_1 = 0$ in large $N$ limit, i.e. the winding in the
Euclidean time direction is a good quantum number. We thus
identify it with the thermal AdS background. $\mu_3$, which is a
minimum, can then be identified with the Euclidean big black hole
phase (BBH). $\mu_2$, which has a unique negative eigenvalue can
be identified with the Euclidean small black hole phase (SBH) in
AdS. That a small black hole in AdS has a unique negative
eigenvalue was pointed out in~\refs{\perry}. Moreover, one can
show from the effective potential \egst\ and \buiop\ that $\mu_2$
has a negative specific heat while $\mu_3$ has a positive specific
heat, without knowing their explicit dependence of $(a,b)$ on $T$.
The derivation is given in Appendix B. This matches well with the
thermodynamic properties of the small and big black holes.

For fixed 't Hooft coupling $\lam$, as one varies the temperature
$T$, $(a (\lam, T), b (\lam,T))$ trace a curve in the $a-b$ plane,
which we will denote by $\CC_\lam$. Any curve $\CC_\lam$ in the
$a-b$ plane which starts below curve I in \contour\ at low enough
temperature and ends up to the right of the vertical line $a=1$ at
sufficiently high temperature (assuming it intersects curves
I,II,III and $H$ only once) reproduces qualitatively the
Hawking-Page picture\foot{We have drawn such a hypothetic curve in
\contour.}. The thermal history following such a $\CC_\lam$ can be
described as follows. At sufficiently low temperature, we start
with some point below curve I, where the theory has a unique
critical point $\mu_1$, corresponding to thermal AdS. As $T$
increases to a temperature $T_0$, $\CC_\lam$ will intersect the
curve I \curvIa, where new critical points $\mu_2$ (SBH) and
$\mu_3$ (BBH) come into existence. At $T_1 > T_0$, it intersects
with curve II, at which $\mu_1$ and $\mu_3$ change dominance. This
is the Hawking-Page transition. Above $T_1$, $\mu_3$ (BBH phase)
dominates and thermal AdS becomes only metastable. As $T$
increases further $\CC_\lam$ will eventually hit $a=1$ (line $H$)
from the left, where the large $N$ expansion (perturbative string
expansion) around $\mu_1$ (thermal AdS) breaks down. This
temperature should be identified with the Hagedorn temperature
$T_H$ in AdS string theory.

There are a few other important features of our matrix model not
visible in the bulk supergravity analysis:

\item{1.} We found that there exists a line III where the SBH
phase undergoes a Gross-Witten transition from a gapped phase to a
gapless one in the eigenvalue distribution of $U$. We will denote
$T_c$ the temperature where $\CC_\lam$ intersects with line III.
Since line III lies between line I and line $a=1$, we should have
$T_0< T_c < T_H$. $T_c$ can be lower or higher than the
Hawking-Page temperature $T_1$ depending on where $\CC_\lam$ hits
line III. This phase transition for SBH, while not visible in
supergravity, has a natural interpretation in string theory. We
would like to identify it with the so-called Horowitz-Polchinski
correspondence point~\refs{\horoP} for SBH, i.e. the point at
which the horizon size of SBH is comparable to the string
scale\foot{Note that the SBH should be considered as a
ten-dimensional black hole.}. In~\refs{\horoP} it was argued that
as one adiabatically decreases the string coupling, a black hole
makes a transition to a state of highly excited strings with the
same quantum numbers (such as mass, charge, angular momentum etc).
Here we fix the string coupling (i.e. $N$), but raise the
temperature adiabatically. The horizon radius of a small black
hole decreases and eventually reaches the string scale.
Following~\refs{\horoP} we would like to argue that beyond $T_c$,
it is more appropriate to view the critical point \theyo\ as
describing a set of highly excited string states. That the phase
transition is third order suggests that the energy, entropy and
specific heat of a SBH vary continuously across the correspondence
point as it becomes a highly excited string states, but
derivatives of the specific heat jump in the large $N$ limit. In
the limit of large $\lam$, i.e. $R \gg l_s$, we expect that $T_c$
should be close to $T_H$ and much greater than the Hawking-Page
temperature $T_1$. That $T_c$ is below $T_H$ appears to be
consistent with the physical picture of the microcanonical
ensemble (see e.g.~\refs{\barbon,\aharony}). For notational
simplicity, below we will continue to refer to \theyo\ as the SBH
phase, keeping in mind that it should really correspond to a
highly excited string state.

\item{2.} Our matrix model indicates that at the Hagedorn
temperature $T_H$, the critical points associated with thermal AdS
and the SBH merge together. This appears natural since very close
to the Hagedorn temperature, thermal AdS will be dominated by a
few long string states. One expects that the distinction between
thermal AdS and the highly excited string phase which the SBH
becomes above $T_c$ disappears at the Hagedorn temperature. This
is again consistent with the physical picture of the
microcanonical ensemble of~\refs{\barbon,\aharony}.

\item{3.} Above $T_H$, i.e. when the $\CC_\lam$ curve goes to the
right of line $a=1$, thermal AdS becomes tachyonic and the
critical point corresponding to SBH disappears. BBH remains the
only stable phase. The physical interpretation of a tachyonic
thermal AdS is not completely clear to us. In the past, it has
been argued that string theory above the Hagedorn temperature can
be interpreted again as some kind of long string phase which can
be analyzed by analytic continuation\foot{This argument was made
in flat space. When the radius of AdS is much larger than the
string scale, we expect the behavior of thermal AdS to be similar
to that of flat space.} (see e.g.~\refs{\Tan}). Our result does
not contradict this point of view. But we should note that this
tachyonic thermal AdS phase cannot be reached from the
microcanonical ensemble. Whether it plays any role in the
canonical ensemble is not clear to us.

\medskip

We now make some comments on the possible dependence of curve
$\CC_\lam$ (i.e. curve $(a(\lam,T), b(\lam,T))$ with $\lam$ fixed)
on the 't Hooft coupling $\lam$. At $\lam=0$, i.e. free theory,
$b(T)=0$~\refs{\sundborg,\aharony}. $\CC_0$ moves along the
$a$-axis from $a=0$ at $T=0$ to $a \to \infty$ at $T \to \infty$.
It crosses all lines in \contour\ at a single point $(a=1,b=0)$.
This was the case analyzed in~\refs{\LiuVY}. At weak coupling it
has been shown in~\refs{\minW} that for pure gauge theory at weak
coupling, $b(\lam,T) = O(\lam^2) > 0$ and $a(\lam,T)=a(0,T)+
O(\lam^2)$. If the result also holds for $\NN=4$ SYM theory, then,
$\CC_{\lam\ll 1}$ corresponds to a curve slightly rising above the
horizontal $a-$axis. In this case $T_c < T_1$. In the supergravity
limit $\lam \gg 1$, $T_0, T_1 \sim {1 \ov R}$ while $T_c, T_H \sim
{1 \ov l_s}$, i.e. there is a big hierarchy between these
temperature scales. One can in principle determine part of
$\CC_\lam$ for $\lam \rightarrow \infty$ by equating the free
energy in the gauge theory with the corresponding free energy in
supergravity. To leading order in large $N$ we equate the
Euclidean actions of the SBH and BBH to the corresponding actions
of $\mu_2$ and $\mu_3$ in the gauge theory. More explicitly,
\eqn\freeAC { Q(\mu_2)={I_{-} \ov N^2}, \qquad Q (\mu_3)={I_{+}
\ov N^2} } where $I_{-}$ and $I_{+}$ are the Euclidean actions
(given in \exacR) of the SBH and BBH respectively and $Q$ is given
by \egst. Since $I_{-}$ and $I_{+}$ are functions of the single
variable $t=T/T_0$, we can use \freeAC\ to determine $a(t)$ and
$b(t)$ in the $\lam \rightarrow \infty$ limit. Note that this
comparison is only valid between curve I and curve III, since
above curve III, $\mu_2$ undergoes a large $N$ phase transition
which is not visible in supergravity.

To summarize, our phenomenological $(a,b)$ model reproduces all
the important features of string theory in $AdS_5 \times S_5$ at
finite temperature. In fact we got more. We found a description of
the Horowitz-Polchinski correspondence point in terms of a
Gross-Witten transition and a non-perturbative picture at and
beyond the Hagedorn temperature for thermal AdS.

After finding the critical points one can then use \furmo\ to find
the large $N$ expansion of $\ZZ(a,b)$ around them. These
expansions should correspond to perturbative string expansions
around the corresponding bulk geometries. In the rest of the paper
we will examine in detail various regions of the parameter space
where these expansions break down and study the physics there.
Such regions include:

\item{1.} At $T_0$, where the BBH and SBH saddles merge together.
Perturbative string expansions around BBH and SBH are not valid.

\item{2.} At the Hawking-Page temperature $T_1$, where there is a
first order phase transition between thermal AdS and BBH. Although
the large $N$ expansion around each critical point does not break
down, the large $N$ expansion of the full partition function
requires a special treatment.

\item{3.} At $T_c$, where the Gross-Witten transition for SBH
takes place.

\item{4.} At $T_H$, the Hagedorn temperature of thermal AdS.

\ndt Note that in $1$ and $4$ above, the breakdown in the large
$N$ expansion happens for the metastable phases. In the
terminology of the first order phase transitions, $T_0$ and $T_H$
are the spinodal temperatures for the BBH and thermal AdS
respectively, beyond which the metastable phase become unstable.

\subsec{Full partition function and smoothening of Hawking-Page
transition}

An immediate consequence of considering the theory at finite $N$
is that the sharp Hawking-Page transition is smoothed out to a
region of width of order $N^{-2}$.

In the infinite $N$ limit, the partition function of the system is
\eqn\Ifullpar{ \log \ZZ = \cases{\log K_1 + O(1/N^2) & $T < T_1$
\cr\cr -N^2 Q (\mu_3) + \log K_3 + O(1/N^2) & $T > T_1$ } } where
$K_1$ and $K_3$ are Gaussian factors computed from the integral
\yhios. Recall that $Q(\mu_1) =0$. $Q(\mu_3)$ equals to zero at
$T_1$ and become negative (positive) above (below) $T_1$. The
transition is first order with a nonzero latent heat given by
\eqn\lateH{ E = N^2 {\p Q(\mu_3) \ov \p \beta}\bigg|_{T_1} + O(1)
} The expectation value of the Polyakov loop also jumps at $T_1$
\eqn\eigDis{ \rho_1^2 (T) = \cases{O(1/N^2) & $T < T_1$ \cr\cr {1
\ov 4} \le(1 + \sqrt{1-{1 \ov \mu_3}} \ri)^2 & $T > T_1$ \cr } }

At finite $N$ we need to include contributions from both
geometries. The full partition function of the system between
temperature $T_0$ and $T_H$ can then be written in terms of the
following asymptotic expansion \eqn\fullpar{ \ZZ \approx e^{-N^2 Q
(\mu_1)} A_1 + e^{-N^2 Q (\mu_3)} A_3 } where $A_1, A_3$ are
asymptotic series around $\mu_1$ and $\mu_3$ respectively
\eqn\oneser{\eqalign{ A_1 & = K_1 \le (1 + \sum_{n=1}^\infty
N^{-2n} c_n (a,b) \ri)\cr
A_3 & = K_3 \le (1 + \sum_{n=1}^\infty N^{-2n} d_n (a,b) \ri)\cr
}} Note that there is no contribution from $\mu_2$, which is a
maximum of $Q(\mu)$. Below the Hawking-Page temperature $T_1$, the
$\mu_1$ saddle dominates and $\mu_3$ is only metastable. The
contribution of the second term in \fullpar\ is exponentially
small compared with the first. Their roles reverse above $T_1$.

The sharp transition at $T_1$ in smoothened out at finite $N$ into
a finite region $T - T_1 \sim O(N^{-2})$. We now examine this
cross over region in some detail. At ${T - T_1 \ov T_1} = \ep \ll
1$, we can expand $Q(\mu_3)$ as \eqn\thgp{ Q (\mu_3(a(T),b(T)),
a(T), b(T)) = - \nu \ep + O(\ep^2), \qquad
} where \eqn\defgT{ \nu = \rho^2 p_1 + \rho^4 q_1 > 0 , \qquad p_1
= T_1 {\p a \ov \p T} \biggr|_{T_1}, \qquad q_1 = T_1 {\p b \ov \p
T} \biggr|_{T_1} } and $ \rho^2 = \rho_1^2 (\mu_3 (T_1)) $. Note
the latent heat \lateH\ is related to $\nu$ by $ E = N^2 \nu T_1
$.

To focus on the transition region we consider the following limit
$$
N \to \infty, \qquad \ep \to 0, \qquad t = \ep N^2 = {\rm finite}
$$
In this limit we find that \eqn\criex{ \log \ZZ = \log \le(K_1
(T_1) + K_3 (T_1) e^{\nu t} \ri) + O(N^{-2}) } \criex\ smoothly
interpolates between the first and second line of \Ifullpar\ as
$t$ varies from $-\infty$ to $+\infty$. The expectation value of
the Polyakov loop is given by \eqn\finr{ \rho_1^2 (t) = {1 \ov
N^2} {\p \log Z \ov \p a}= \rho^2 {K_3 (T_1) e^{\nu t} \ov K_1
(T_1) + K_3 (T_1) e^{\nu t}} + O(1/N^2) } $\rho_1^2$ smoothly
interpolates between the first and the second line of \eigDis.

\newsec{SBH and tunnelling}

As pointed out in~\refs{\Hawkingpage} when $T > T_0$\foot{In
contrast, the flat space has a non-perturbative instability at any
finite temperature~\refs{\GrossPY}.} thermal AdS and BBH can
tunnel into each other with the Euclidean SBH as the instanton
bounce~\refs{\GrossPY}. Our effective potential $Q(\mu)$ (see
\Qshape) gives a concrete realization of the physical picture.
Through thermal fluctuations, BBH or thermal AdS can jump to the
top of the barrier to become a SBH. Since it has negative specific
heat, the small black hole then can either become thermal AdS by
emitting thermal Hawking radiation or become a big black hole by
absorbing radiation. In the Euclidean description, this
corresponds to rolling down the hill from the two sides of the
effective potential. In thermal equilibrium, of course the total
probability to go from thermal AdS to BBH or vice versa should be
the same.

\subsec{Tunnelling between thermal AdS and BBH}

In this subsection we will calculate the tunnelling rates between
thermal AdS and BBH using our effective potential description. For
definiteness, we will restrict to the temperature range $T_{c}> T
T_{0}$ (i.e for $(a,b)$ lying between line I and line III). In
this range the details of smoothening of singular behavior of the
effective potential \DFQ\ at $\mu=1$ by non-perturbative effects
will not be relevant and we will use \DFQ\ to perform the
analysis. The discussion for the range $T_{H}>T>T_{c} $ is
similar, and will not be repeated. The only difference is that
since the SBH is precisely located at $\mu=1$ for $T_{H}>T>T_{c}$,
it is therefore more convenient to use the two-dimensional
effective potential~\defVg. The perturbative expansion around SBH
breaks down near $T_c$, and will be discused in next subsection.

The tunnelling rates between thermal AdS and BBH can be readily
computed using the effective potential following the standard
procedure~\refs{\langer,\callancoleman,\affleck}. To be definite,
let us first consider the tunnelling rate for thermal AdS over the
barrier. In computing \furmo, instead of using the $\mu$ contour
going from $-\infty$ to $+\infty$ along the real axis, we consider
a contour $C_1$ along the real axis from $-\infty$ to $\mu_2$ and
then deform the contour at $\mu_2$ along a steepest descent
contour to the complex $\mu$ plane (see Fig.~3a). The partition
function obtained using contour $C_1$ is given by \eqn\parAo{
\ZZ_1 \approx e^{-N^2 Q (\mu_1)} K_1 \le(1 + O(N^{-2}) \ri) + {i
\ov 2} e^{-N^2 Q (\mu_2)} K_2 \le(1 + O(N^{-2}) \ri) } where $K_1$
and $K_2$ arise from the Gaussian factor in the saddle point
approximation. The resulting free energy (let us call it $F_1$)
has an imaginary part given by \eqn\ImFi{ {\rm Im} F_1 \approx {1
\ov 2 \beta} e^{-N^2 (Q(\mu_2) - Q(\mu_1))} {K_2 \ov K_1} \le(1 +
O(N^{-2}) \ri) } The tunnelling rate is then obtained from \ImFi\
by~\refs{\affleck} \eqn\tranRi{ \Ga_1 \approx {\om_0 \beta \ov
\pi} {\rm Im} F_1 = {\om_0 \ov 2 \pi} e^{-N^2 (Q(\mu_2) -
Q(\mu_1))} {K_2 \ov K_1} \le(1 + O(N^{-2}) \ri) } where $\om_0$ is
the frequency for the unstable mode around the SBH background.
Similarly the tunnelling rate from BBH to thermal AdS can be
obtained by computing \furmo\ along a contour $C_2$ (see Fig. 3b)
to be \eqn\thnn{ \Ga_2 \approx {\om_0 \ov 2 \pi} e^{-N^2 (Q(\mu_2)
- Q(\mu_3))} {K_2 \ov K_3} \le(1 + O(N^{-2}) \ri) }

\ifig\newfig{(a) plots the contour $C_1$ which is used to compute
\ImFi\ and (b) plots the contour $C_2$ used to compute \thnn.}
{\epsfxsize=12cm \epsfbox{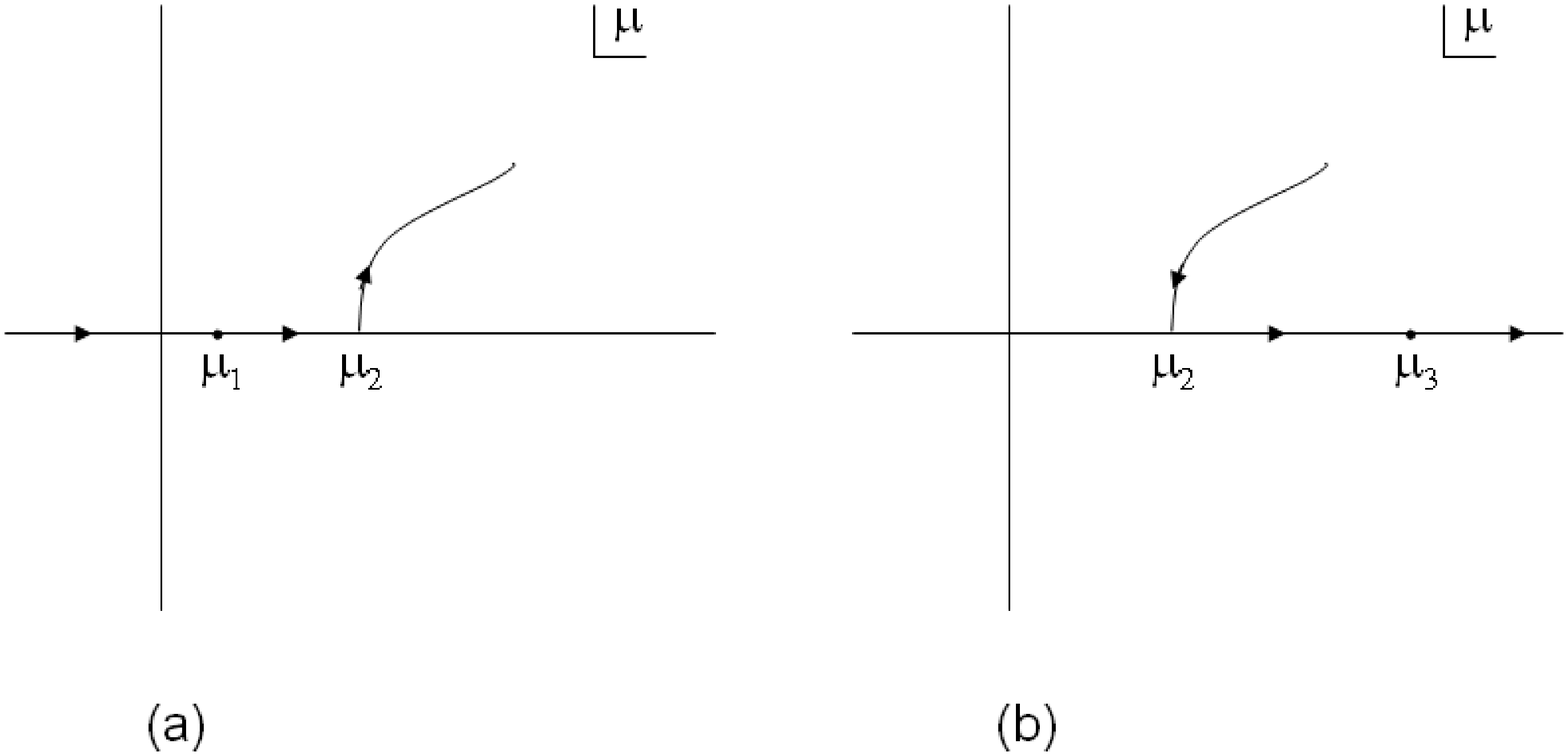}}

Note that the imaginary part of the partition functions with
contours $C_1,C_2$ can also be obtained by a Borel resummation
over the perturbative expansion around the BBH and thermal AdS
respectively. For example, the asymptotic expansion around the BBH
$\mu_3$ has the form \eqn\bbhs{ e^{-N^2 Q (\mu_3)} K_3 \le (1 +
\sum_{n=1}^\infty N^{-2n} d_n (a,b) \ri) } Due to the presence of
a maximum at $\mu_2$, the coefficients $d_n$ at large $n$ are
given by \eqn\largBe{ d_n \approx \frac{K_{2}}{2\pi K_{3}}{ \Ga(n)
\ov (Q(\mu_{2})- Q(\mu_{3}))^{n}} } The asymptotic expansion
\bbhs\ is clearly divergent and not Borel summable. In fact the
Borel transform $\sum_{n}\frac{d_{n} z^{n}}{n!}$ of $\sum_{n}
\frac {d_{n}}{N^{2n}}$ is singular at $z= Q(\mu_{2}) - Q(\mu_3)$,
thus preventing Borel summability if the difference between the
action of SBH and BBH $Q(\mu_{2}) - Q(\mu_3) $ is positive. This
is a familiar situation in instanton physics, where we can
interpret the singularity in the Borel transform in the positive
axis as real instantons. One can integrate the Borel transform
just above the singularity on the real positive axis, producing an
imaginary part. Using this procedure, we find that \bbhs\
becomes\foot{One keeps only finite number of terms in the first
term below. Also there is a sign ambiguity here. We take the sign
to be the same as that of contour $C_2$.} \eqn\thsn{ e^{-N^2 Q
(\mu_3)} K_3 \le (1 + O(N^{-2}) \ri) - {i \ov 2} e^{-N^2 Q
(\mu_2)} K_2 \le(1 + O(N^{-2}) \ri) } which is precisely what we
get by computing the partition function following contour $C_2$ in
Fig. 3b and leads to the tunnelling rate \thnn. Similar
discussions apply to the perturbative expansion around thermal
AdS. Note that the instanton effect we have obtained associated
with the SBH is of order $e^ {-\frac{1}{g_{s}^{2}}}$ for $g_{s} =
\frac{1}{N}$ and not of order $e^{-\frac{1} {g_{s}}}$ typical of
D-instantons. In some sense, the effect we are describing could be
interpreted as a collective state of $N$ D-instantons.

Of course the total partition function is real and therefore
cannot contain any imaginary part. Physically this is equivalent
to a balance rule between the probability to nucleate a SBH from
thermal AdS and the probability of decay of the metastable BBH.
The two imaginary parts that we obtain in contours $C_1$ and $C_2$
should cancel in the complete partition function that is real.
This is in fact the case once we realize that the sum of contours
$C_1$ and $C_2$ gives the real contour which defined the total
partition function $\ZZ$.

To conclude this subsection we will briefly comment on the
asymptotic expansion for $T<T_{0}$. Below line I the two saddles
$\mu_{2}$ and $\mu_{3}$ move into the complex plane, in other
words the real instanton that we have associated with the SBH
above line I becomes complex. This in particular means that the
corresponding singularity in the Borel transform is not any more
on the real positive axis and Borel summability is potentially
restored\foot{This phenomena is well known in quantum field
theory. For instance for the $\lambda \phi^{4}$ theory we have
complex instanton solutions that produce a priori harmless
singularities in the Borel transform on the negative real axis.}.
{}From the physical point of view this means that the thermal AdS
saddle is stable below line I.

\subsec{A Gross-Witten transition for small black hole}

In this subsection we examine in detail the behavior of SBH as it
crosses line III ($c=1$) in the $a-b$ plane. It undergoes a third
order phase transition in the large $N$ limit and the large $N$
expansion around SBH breaks down there. We will define a double
scaling limit to smooth out the transition.

When $c = 1 + \ep$ with $0 < \ep \ll 1$, we find from
\solui--\critone\ that the critical point corresponding to the SBH
is given by \eqn\hdts{ \mu_2 = 1 + {\ep^2 \ov 4} + \cdots , \qquad
g_2 = 1 + {\ep \ov 2} + \cdots, \qquad \rho_1 (g_2) = \ha + {\ep
\ov 4} + \cdots } When $c=1 - \ep$, one finds from \theyo\ that
\eqn\hdtn{ \mu_2 = 1 , \qquad g_2 = 1 - {\ep \ov 2} + \cdots,
\qquad \rho_1 (g_2) = \ha - {\ep \ov 4} + \cdots } Thus as $c$
crosses $1$, $g_2$ crosses $1$ and the eigenvalue distribution of
$\mu_2$ crosses $\rho_1 = \ha$. One can also check that the second
derivatives around the SBH varies smoothly across $c=1$.

As we commented after equation \DEFw, \defVg\ has a third order
discontinuity at $g=1$ which is smoothened out at finite $N$. More
precisely, let \eqn\sfsg{ g = 1 - N^{-{2 \ov 3}} y \qquad }
Then $V$ should be replaced by\foot{See e.g. sec. 3.1
of~\refs{\LiuVY}. $F_n$ below correspond to $F_n^{(2)}$ there.}
\eqn\fgsV{ V = {1 \ov 4b} (\mu-a)^2 + {g^2 \ov 4}{1-\mu \ov \mu} -
\sum_{n=0}^\infty N^{-{2 \ov 3} n} F_n (y) } where the $F_n (y)$'s
are smooth functions of $y$. In particular, $F_0$ describes the
doubling scaling limit of \Hjho\ with the following asymptotic
expansion \eqn\mquanii{ F_0 (y) = \cases{ {y^3 \ov 6} - {1 \ov 8}
\log (-y) + \cdots & $-y \gg 1$ \cr\cr {1 \ov 2 \pi} e^{- {4
\sqrt{2} \ov 3} y^{3 \ov 2}} \le( - {1 \ov 8 \sqrt{2} y ^{3 \ov 2}
} + \cdots \ri) & $ y \gg 1$ \cr }} Note that \fgsV\ smoothly
interpolates between the second and the third line of \defVg.

Equations \hdts--\fgsV\ suggest that to study the physics of the
SBH near $c=1$, we can consider the following scaling \eqn\onesc{
a(T) = a_0 + a_1 \ep q , \qquad b (T) = b_0 + b_1 q \ep, \qquad {2
(1-a_0) \ov b_0} =1 } \eqn\secS{ \mu = 1 + \ep^2 x , \qquad g = 1
- \ep y, \qquad \ep = N^{-{2 \ov 3}} } where $a_0 = a (T_c), a_1 =
T_c a'(T_c)$ (similar for $b$) and $\ep q = {T- T_c \ov T_c}$.
Thus we have
$$
c ={2 (1-a) \ov b} = 1 - c_1 \ep q , \qquad c_1 = {a_1 \ov 1-a_0}
+ {b_1 \ov b_0} ={1 \ov b_0} (2a_1 + b_1)
$$
Note $c_1>0$ according to our assumptions on the thermal history
of the model and $q<0$ as we approach line III form below.

Plugging \onesc\ and \secS\ into \fgsV, we find that \eqn\uynp{
N^2 V={N^2 (1-a)^2 \ov 4b} + \ha x \le(y- {c_1 q \ov 2} \ri) - F_0
(y) + O({\ep }) } Note that in the last expression, $x$ simply
plays the role of a Lagrange multiplier. The integral over $x,y$
in \yhios\ is not well defined since we are integrating around the
neighborhood of a saddle point in the $x-y$ plane. If we rotate
the integration contour of $x$ to be along the imaginary axis, the
$x$ integral will result in a delta function for $y$ and we find
that the partition function around the SBH is given by \eqn\value{
\ZZ_{SBH} = i N \sqrt{\pi \ov b} e^{-{N^2 (1-a)^2 \ov 4b} + F_0
\le({c_1 q \ov 2}\ri) } \le (1 + O(\ep) \ri) } The prefator $i$
can be understood as due to the tachyonic mode of the SBH. We find
that around $T_c$, one can define a double scaling limit where the
SBH is described by $F_0$. It was argued in~\refs{\kms} that $F_0
(t)$ describes the full partition function of the type 0B theory
in $d=0$ dimension, i.e. pure 2-d supergravity. The parameter $t$
is proportional to the cosmological constant $\mu$ in the
super-Liouville interaction. We are then led to the conclusion
that in a double scaling limit around $T_c$ (as we argued earlier
in the Horowitz-Polchinski correspondence point) the SBH appears
to be described by type 0B theory in zero dimension.

\newsec{Catastrophes and the break down of perturbative string
expansions}

In this section we examine the critical behavior of the BBH at
$T_0$ where its saddle merges with that of the SBH and the
critical behavior of the thermal AdS at the Hagedorn temperature
$T_H$. We find in both places that the breakdown of the large $N$
expansion can be understood in terms of the simplest type of
catastrophes allowed by the symmetry. The divergences at the
perturbative level can be smoothened out at finite $N$ using the
standard techniques of catastrophe theory.

\subsec{Nucleation of black holes}

The large $N$ expansion around the big black hole saddle breaks
down near line I, where it coalesces with the unstable small black
hole saddle. We will show that the critical behavior there is
given by the fold catastrophe. One can define a double scaling
limit in which the partition function for this sector is given by
an Airy function.

{}From \curvIa, curve I can be parameterized by
\eqn\curvIaa{\eqalign{ a(w) & = {1 - 2 w \ov (1- w)^2 (1 + w)},
\cr b (w) & = {2 w \ov (1- w)^2 (1 + w)^3}, \qquad w \in [0,1] \ .
}} Suppose that at temperature $T_0$, the curve $(a(T), b(T))$
intersects the curve I \curvIaa\ at a point labelled by $w_0$,
i.e. $ a(T_0) = a(w_0), \; b(T_0) = b(w_0)$. At the intersection
point, \critone\ has a double root given by $w_0$, which is an
inflection point of $Q (w)$. We will consider the behavior of $Q
(w)$ \egst\ near $w = w_0$ and $T=T_0$. Let \eqn\Imparm{ a (T)= a
(T_0) + a_1 \ep , \qquad b (T)= b (T_0) + b_1 \ep , \qquad w = w_0
+ y } with $\ep = {T-T_0 \ov T_0}$ and $a_1 = T_0 a'(T_0), \; b_1
=T_0 b' (T_0)$. We will consider the regime $|\ep| \ll 1$ and $|y|
\sim \sqrt{|\ep|} \ll 1$. Plugging \Imparm\ into \egst\ and
expanding in $\ep$ and $y$ we find that\foot{The expansion below
breaks down at $w_0 = 0$, where it can be checked that $C_0$ and
various higher order terms become singular. At $w_0 =0$, we have
$a_0 = 1, \; b_0 = 0$ and the physics goes over to that
of~\refs{\LiuVY}. } \eqn\simFQ{ Q = C_0 (\ep,w_0) - f \le(-{1 \ov
3} y^3 + q \ep y\ri) + O(\ep^2, y^4, y^2 \ep) } with \eqn\gqde{ f
= -\ha a'(w_0) , \qquad q = {(1+w_0) a_1 \ov b'(w_0)} \le( \tan
\th_0 - {b_1 \ov a_1} \ri) } where $\tan \th_0$ is the slope of
line I at $w_0$. Note that
$$\eqalign{
& a'(w_0) < 0 , \qquad b'(w_0) > 0, \qquad \tan \theta_0 =
{b'(w_0) \ov a'(w_0)} = -{2 \ov (1+w_0)^2}, \qquad \theta_0 \in
(\pi/2, \pi) }
$$
{}From our assumption of the thermal history of the theory, $ q$
is positive. Note that $C_0 (\ep, w_0)$ is analytic in $\ep$.

For $\ep > 0$, from \simFQ\ $Q$ has one maximum and one minimum at
$y = \pm \sqrt{ \ep q}$. The two extrema merge at $q=0$ and move
to complex values for $\ep < 0$. The values of $Q$ at the minimum
and the maximum are given by ($\ep>0$) \eqn\resuL{ Q_0 = C_0 (\ep)
\mp { 2 f q^{3 \ov 2} \ov 3 } \ep^{3 \ov 2} + O(\ep^2) } The
second term gives the leading nonanalytic term in $\ep$ and the
specific heat has a critical exponent $\ga = \ha$.

It is instructive to compare \resuL\ with the result \exacR\ from
supergravity. Expanding \exacR\ in $\ep$ around $\beta_0$ for big
and small black holes we find \eqn\sugH{ I = \tilde C_0 (\ep) \mp
{2 } \ep^{3 \ov 2} + \cdots } with $\tilde C_0$ an analytic
function of $\ep$. We see exact agreement between \resuL\ and
\sugH\ in the critical exponent. The critical exponent $\ha$ is
universal, depending only on the fact that a maximum and a minimum
merge together (fold catastrophe).

As $T \to T_0$, the large $N$ expansion around the big and small
black holes break down. The physics around them can be captured by
a double scaling limit. Since we are interested only in the BBH
and SBH, we will again consider \furmo\ along contour $C_2$ in
Fig. 3b. From \simFQ\ we introduce a new variable
$$
z = q \ep (N^2 f )^{2 \ov 3}
$$
and consider the scaling limit
$$
\ep \to 0, \qquad N \to \infty, \qquad z = {\rm finite}
$$
In this limit, the partition function becomes \eqn\finR{ \ZZ_2 =
\CC_0 \int_{C_2} ds \, e^{-{1 \ov 3} s^3 + z s } \ } which is
given by an Airy function. $C_0$ is a non-universal factor
$$
\CC_0 = (N^2 f)^{-{1 \ov 3}} {2w_0 \ov (1-w_0^2)^2} e^{\FF_1
(w_0)} e^{-N^2 C_0}
$$
where $\FF_1$ is the $O(1/N^2)$ term in $\FF(\mu)$ \resff\ (not
given explicitly there). It is easy to check that our choice of
contour gives the Airy function \eqn\Gepoa{ \ZZ_2 = 2 \pi i \,
\CC_0 e^{-{2 \pi i \ov 3}} Ai (z e^{-{2 \pi i \ov 3}}) } \Gepoa\
smoothes out the divergences in perturbative expansions. Also note
that the argument of the Airy functions precisely sits on the
Stokes line. This is a consequence of the fact that for $\mu_2,
\mu_3$ real, ${\rm Im} Q (\mu_2) = {\rm Im} Q (\mu_3) = 0$.
Although \Gepoa\ is complex, the full partition function should be
real when including the contribution from the contour $C_1$ of
fig. 3a.\foot{We would like to thank G.~Festuccia and
A.~Scardicchio for extensive discussions regarding this point.}

\subsec{Hagedorn behavior for thermal AdS}

We will now examine the merger of thermal AdS and the SBH (or more
precisely long string phase). For this purpose, let us first look
at the free energy near the thermal AdS background for $a<1$,
which can be found by expanding the partition function around the
saddle \theads. Around \theads, the partition function can be
approximated as \eqn\resuO{\eqalign{ \ZZ_1 & = {N \ov 2 \sqrt{\pi
b}} \int_{-\infty}^1 d \mu \, e^{-{N^2 \ov 4b } (\mu -a)^2 } \, {1
\ov (1-a) - (\mu -a)} \cr & \approx {N \ov 2 \sqrt{\pi b}}
\sum_{n=0}^\infty \int_{-\infty}^\infty d x \, e^{-{N^2 \ov 4b}
x^2}\, {x^n \ov (1-a)^{n+1}} \cr & = {1 \ov 1-a} \sum_{n=0}^\infty
{\Ga(n + \ha) \ov \sqrt{\pi}} \le({2 \sqrt{b} \ov N (1-a)}
\ri)^{2n} \cr }} There are also nonperturbative corrections which
are omitted here. Thus the free energy around the thermal AdS
background can be written as \eqn\freeEnA{\eqalign{ \log \ZZ_1 & =
- \log (1-a) + {2 b \ov (1-a)^2 N^2} + +{10 b^2 \ov (1-a)^4 N^4} +
{296 b^3 \ov 3 (1-a)^6 N^6} + \cdots \cr }} Note that in contrast
to the free theory case analyzed in~\refs{\LiuVY}, the free energy
now receives perturbative contributions to all orders. The free
energy diverges as $a \to 1$. The leading order divergence,
arising from genus one contribution, is \eqn\ahgeD{ \log \ZZ_1 = -
\log (T - T_H) + {\rm const} } since as $a \to 1$, $1-a \propto
T-T_H$. \ahgeD\ is precisely the Hagedorn divergence for string
theory in a spacetime with all directions compactified (recall
that AdS behaves like a box). By a Laplace transform of \ahgeD\
one finds that the density of states is given by \eqn\lowtEn{ \Om
(E) \approx {\rm const} \; e^{\beta_H E} \le(1+ O(1/E^2) \ri) }
Also note that as $T \to T_H$, the free energy around SBH is given
by \eqn\freeSBH{\eqalign{ \log \ZZ_1 \propto - N^2 (T-T_H)^2 +
\cdots }}

The perturbative expansion \freeEnA\ breaks down at $1-a \sim
N^{-1}$. To explore the physics near this point, we let\foot{Below
for notational simplicity we will assume $b$ does not change as
$a=1$ is crossed. To incorporate the change in $b$ is
straightforward.} \eqn\parade{ a = 1 - N^{-1} q , \qquad \mu = 1 -
N^{-1} x, \qquad g = 2 N^{-\ha} y } and consider a double scaling
limit with $N \to \infty$ with $q,x,y$ finite. We find that
$$
Q = N^{-2} P + O(N^{-3})
$$
with \eqn\inteP{ P = {(x-q)^2 \ov 4 b} + {xy^2 } \ . } Note that
$P$ has two critical points for $q > 0$
$$\eqalign{
& (1) \qquad x_1=q, \qquad y_1 = 0, \qquad P'' = \pmatrix{{1 \ov 2
b} & 0 \cr 0 & 2q \cr} \cr & (2) \qquad x_2=0, \qquad y_2 = \sqrt{
q \ov 2b}, \qquad P'' = \pmatrix{{1 \ov 2 b} & \sqrt{2 q \ov b}
\cr \sqrt{2 q \ov b} & 0 \cr} \cr }$$ corresponding to $(\mu_1,
g_1)$ and $(\mu_2, g_2)$. They merge at $q=0$ and for $q < 0$ only
the first solution remains. The relevant part of the partition
function \yhios\ then becomes
 \eqn\parGh{\eqalign{
  \ZZ_1 & = {N
\ov \sqrt{\pi b}} \int_{-\infty}^\infty dx \, \int_0^\infty y dy
\, e^{- P (x,y)} \cr & = 2 N \int y dy \, e^{- q y^2 + b y^4} \cr
 }}

Introducing a (0-dimensional) complex scalar $\phi$ with $y=
|\phi|$, the second line of \parGh\ can also be written as
 \eqn\newWex{
 \ZZ_1 = \int d \phi d \phi^* \, \exp \le[ - m^2
 (\beta) \phi^* \phi + b (\phi \phi^*)^2 \ri]
 }
 with $m^2 (\beta) = q \propto T-T_H$.
 It is tempting to identify $\phi$ with the
so-called thermal scalar (a winding tachyon) in string
theory\foot{Note that we see a zero dimensional scalar since AdS
may be considered as a box.}. Indeed  \newWex\ coincides with the
effective action one expects for a thermal scalar near Hagedorn
temperature~\refs{\atic}. It is clear from the second line of
\parGh\ or \newWex\ that the merger of thermal AdS and SBH is
described by a cusp catastrophe with $q=0$ (i.e. $a=1$)
corresponding to the cusp point. Note that the appearance of
effective action \newWex\ is forced on us by the $U(1)$ symmetry
of the complex scalar field $\phi$, which corresponds to the $Z_N$
symmetry of the boundary effective action in the large $N$ limit.
The cusp catastrophe is the simplest possibility consistent with
this symmetry. The only nontrivial dynamical input in \newWex\ is
that $b >0$, following from the existence of a first order
Hawking-Page transition.

The integral in \parGh\ is not bounded as is expected since we are
focusing in the neighborhood of the effective potential containing
only thermal AdS and SBH. To define the integral in
\parGh\ we will choose an integration contour analogous to $C_1$
of Fig. 3a in section 5. We take the contour in the $y$-plane to
go from $y=0$ to the maximum $y_2 = \sqrt{q \ov 2b}$ along the
real axis and then go straight up to the complex value at $y_2$.
The integration along the real axis will give us an error
function, which smoothes out the divergences of the perturbative
expansion at $q=0$.

\newsec{Conclusions and discussions}

In this paper we introduced a class of phenomenological models to
understand string theory in $AdS_5 \times S_5$ in the canonical
ensemble. Our models reproduces all the known qualitative features
of the theory. They also have some interesting predictions
including the existence of a third order phase transition for SBH,
which we identify with the Horowitz-Polchinski point. We studied
the simplest model \themodel\ in great detail. We found the
Hagedorn behavior of thermal AdS at $T_H$ and the critical
behavior of nucleation of Euclidean SBH and BBH at $T_0$ are
governed respectively by cusp ($A_3$) and fold ($A_2$)
catastrophe. It is clear these features persist for all models in
the class due to universality of the catastrophe. We believe this
gives strong indication that they capture qualitative behaviors of
a weakly coupled string theory in a large AdS spacetime. Since for
a large radius AdS, the SBH resembles a ten dimensional
Schwarzschild black hole in flat spacetime, and the Hagedorn
behavior for strings in AdS resembles that of flat space, we
expect that the behaviors we observe here may yield clues to
answers for similar questions in flat spacetime.

There are many other questions which can be explored along the
lines of our investigation. For example, it would be nice to have
a better understanding of our proposal that the
Horowitz-Polchinski point for a small black hole should correspond
to a Gross-Witten transition in the boundary theory. It would be
interesting to understand whether the process involves changing
the spacetime topology. Also, since the saddles corresponding to
thermal AdS and SBH merge at $T_H$, we expect that in the
worldsheet sigma model of thermal AdS, turning on the marginal
operator corresponding to the thermal scalar at $T_H$, the theory
can be deformed into a SBH background. It would also be
interesting to understand from the worldsheet point of view what
happens when the tachyon in thermal AdS above the Hagedorn
temperature condenses\foot{It appears clear from the effective
potential picture that the theory will flow to the BBH
background.}. The discussion of~\refs{\eva} might be useful for
this purpose.

We believe that the phenomenological approach developed here can
have many other applications. For example, it would be interesting
to develop an effective potential approach for the tunnelling
discussed in~\refs{\vijay}. It would also be interesting to see
whether one can use our methods to address the problem of black
hole information loss.

\bigskip
\noindent{\bf Acknowledgments}

We would like to thank P.~ Basu, E.~ Brezin, A.~ Dhar,
G.~Festuccia, J.~Frohlich, A.~Jevicki, M.~Luscher,
M.~Marino-Beiras, J.~Polchinski, A.~Scardicchio, S.~ Trivedi,
B.~Zwiebach and especially N.~Kumar, S.~Minwalla and N.~Seiberg
for very useful discussions. We also would like to thank
N.~Seiberg for collaboration at early stages of the project. CG is
partially supported by Plan Nacional de Altas Energias
FBA-2003-02-877. HL is supported in part by Alfred Sloan
Foundation and by funds provided by the U.S. Department of Energy
(D.O.E) under cooperative research agreement \#DF-FC02-94ER40818.
SW would like to thank the Theory Division of CERN for
extraordinary hospitality during a sabbatical year when most of
this work was done. He would also like to thank the KITP Santa
Barbara for hospitality where part of this work was done.

\appendix{A}{Large $N$ phase structure and thermal history for general matrix model
\uhgm}

For completeness, in this section we discuss the phase structure
and thermal history of matrix model \uhgm\ using the Hartree-Fock
approximation.
 The Hartree-Fock treatment of double
trace operators was earlier discussed in~\refs{\das}. The
Hartree-Fock approximation gives equations of motion which can be
solved to find the critical points of the theory and the value of
the action evaluated at the critical points. However from this method
one cannot find the off-shell effective potential (essential for
our purposes).

In the infinite $N$ limit, it is convenient to introduce the
density of eigenvalues
 \eqn\chCor{ \rho (\theta) = {1 \ov N}
 \sum_{i=1}^N \delta(\theta - \th_i) , \qquad - \pi \leq \th < \pi
 }
 with
 $${1 \ov N} \Tr U^n =\rho_n = \int_{-\pi}^\pi {d \th } \,
\rho (\th) \,e^{i n \th}
 $$
\uhgm\ can be written as
 \eqn\patI{
  \ZZ = \int [D \rho] \, e^{-
 N^2 V [\rho]}
 }
 where $V[\rho]$ has the form
 \eqn\Impote{\eqalign{
 V[\rho] & = -\ha \int d \th d \phi \, \rho(\th) \rho (\phi) \, P
\log \le(2 \sin {\th - \phi \ov 2} \ri)^2 - S(|\rho_1|^2) \cr
 }}

Since the potential is symmetric, we can take $\rho_1$ to be real.
The equations of motion following from \Impote\ can be written as
 \eqn\sadle{ \int d \phi \, \rho (\phi) \,
 \cot {\th -\phi \ov 2} = \kappa 
 \sin \th
 }
  with
$$
\kappa = 2 S'(x) \rho_1 \ , \quad x = \rho_1^2 \ .
$$
The solutions to the above equation are well
known~\refs{\gw,\wadiapre,\wadia}, leading to the self-consistent
equations for $\rho_1$ (using \tyiD\ with $g$ replaced by $\kappa$
above)
 \eqn\seldcon{\eqalign{
 & \rho_1 = S'(x) \rho_1 , \qquad 0 \leq \rho_1 \leq \ha \cr
 & S'(x) ={1\over 4\rt (1-\rt)}, \qquad \ha \leq
\rho_1 \leq 1 \cr
 }}
Note that the first line of \seldcon\ implies that
 \eqn\phaso{
 \rho_1 = 0
 }
 or
 \eqn\anpoP{
 S'(x)  =1, \qquad x = \rho_1^2 \in [0, {1 / 4}]
 }
We can slightly rewrite the second equation of \seldcon\ as
 \eqn\imphaE{
 S'(x) = f(x), \qquad f(x) = {1 \ov 4 \sqrt{x} (1-\sqrt{x})},
 \qquad x \in [{1 / 4},1 ]
 }

Note $S(x)$ are also functions of 't Hooft coupling $\lam$ and
temperature $T$. We will show that given the following assumptions
about $S(x;\lam,T)$,  \patI\ has exactly the same large $N$ phase
structure as that of the $(a,b)$ model analyzed in the main text
and thus that of the AdS supergravity:
 \item{1.} $S(x)$ is convex, i.e. $S'(x)$ is monotonically
 increasing;
 \item{2.} $S'(x)$ is concave;
 \item{3.} For sufficiently low temperature, $S'(x)$ lies below
 $f(x)$ defined in \imphaE\ in $x \in [{1 / 4},1 ]$. For
 sufficiently high temperature $S'(0)>1$;
 \item{4.} $S'(1/4)$ is a monotonically increasing function of $T$.

\ndt We note that for $(a,b)$ model \themodel, conditions $1$ and
$2$ are automatically satisfied. Condition $3$ corresponds to our
assumption that the $\CC_{\lam}$ curve starts below line I of
\contour\ at sufficiently low temperature and ends to the right of
line $H$ at high enough temperature. Condition $4$ makes sure that
$\CC_{\lam}$ intersects all lines in \contour\ only once as $T$ is
varied. Note that conditions $4$ may be further relaxed.

We first note that $f(x)$ is a monotonically increasing and convex
function which takes values
 \eqn\rhdu{
 f(1/4) = 1, \qquad f(1) = + \infty
 }
 At a sufficiently low $T$, condition 3 implies
\imphaE\ has no solution. Since $S'(1/4) < 1$ and $S'$ is
monotonically increasing, equation \anpoP\ also does not have a
solution. The only phase of the system at this temperature is thus
given by \phaso, i.e.
$$
x_1 =0
$$
This is the thermal AdS. It is a minimum of $V$ for $S'(0)<1$. As
we increase the temperature, the curve $S'(x)$ will start
intersecting\foot{This is guaranteed following our assumptions.}
with $f(x)$. The temperature at which they become tangent is
$T_0$, where the critical points corresponding to SBH and BBH
start appearing. Immediately above $T_0$, \imphaE\ will have two
solutions ${1 \ov 4} < x_2 < x_3 < 1$ and from condition 2 it can
only have two solutions. Since $V$ is bounded from below, $x_3$
should be minimum (BBH), while $x_2$ a maximum (SBH). At this
temperature \anpoP\ again does not have any solution since
$S'(1/4) < 1$. At a temperature $T_c > T_0$, when $S'(1/4) =1$,
$x_2 = {1 \ov 4}$ is both a solution of \anpoP\ and \imphaE. When
$T > T_c$, $S'(1/4) > 1$, \imphaE\ only has one solution $x_3$.
$x_2$ moves to the region $x < 1/4$ and becomes a solution to
\anpoP. Convexity of $S(x)$ implies that the solution to \anpoP\
is unique. At a temperature $T_H$, when $S'(0) = 1$, $x_2$ and
$x_1$ coincide. Above $T_H$, $x_2$ no longer exists and $x_1$
becomes tachyonic due to that $S'(0) > 1$. Since at $T_H$, $x_1$
and $x_2$ coincide, we have $V(x_2) = V(x_1)=0$ in the large $N$.
Since $V(x_2) > V(x_3)$, $V(x_3)$ must be smaller than zero at
$T_H$. We thus conclude that there must be a first order
Hawking-Page transition at some temperature $T_0< T_1<T_H$.

To summarize, the general model satisfying the four assumptions
above have exactly the same phase structure as that of the $(a,b)$
model including the Gross-Witten phase transition for SBH and the
merger of SBH and thermal AdS at $T_H$.

\appendix{B}{Specific heat of small and big black holes}

In this appendix, we show that the phases corresponding to $\mu_2$
and $\mu_3$ have negative and positive specific heat respectively.

We first show that \theyo\ has a negative specific heat. From
\buiop, \eqn\unhne{\eqalign{ c_v (\mu_2) & = - N^2 \beta^2 {\p^2 V
\ov \p \beta^2} \cr & = -{N^2 \beta^2 \ov 2b} \le({\p a \ov \p
\beta} +{1-a \ov b} {\p b \ov \p \beta} \ri)^2 < 0 \cr }}

We now look at the specific heat of a solution to equation
\critone. Evaluated at a critical point $\mu_c (\beta)$, the
action is given by $Q(\beta, \mu_c (\beta))$. We first note an
identity \eqn\anniId{ - {d^2 \ov d \beta^2} Q (\beta, \mu_c
(\beta)) = {\le({\p Q^2 (\beta, \mu) \ov \p \mu \p \beta}\ri)^2
\ov {\p^2 Q \ov \p \mu^2}} \biggr|_{\mu_c} - {\p^2 Q \ov \p
\beta^2} \biggr|_{\mu_c} } In deriving the above equation we have
used the equation of motion ${\p Q \ov \p \mu}|_{\mu_c} =0$ and
\eqn\huan{ {\p \mu_c \ov \p \beta} = - {{\p^2 Q \ov \p \mu \p
\beta} \ov {\p^2 Q \ov \p \mu^2}}\biggr|_{\mu =\mu_c} } We note
from \egst\ that \eqn\unipi{\eqalign{ & {\p^2 Q \ov \p \beta^2}
\biggr|_{\mu_c} = {1 \ov 2b} \le({\p a \ov \p \beta} +{\mu_c-a \ov
b} {\p b \ov \p \beta} \ri)^2 \cr & {\p^2 Q \ov \p \beta \p \mu}
\biggr|_{\mu_c} = -{1 \ov 2b} \le({\p a \ov \p \beta} +{\mu_c-a
\ov b} {\p b \ov \p \beta} \ri) \cr }} Thus the specific heat for
$\mu_c$ can be written as \eqn\unhneC{\eqalign{ c_v (\mu_c
(\beta)) & = - N^2 \beta^2 {d^2 \ov d \beta^2} Q (\beta, \mu_c
(\beta)) \cr & = {N^2 \beta^2 \ov 4b^2} \le({\p a \ov \p \beta}
+{\mu_c-a \ov b} {\p b \ov \p \beta} \ri)^2 \le({1 \ov {\p^2 Q \ov
\p \mu^2}}\biggr|_{\mu =\mu_c} - 2b \ri) \cr }} For $\mu_c =
\mu_2$, ${\p^2 Q \ov \p \mu^2}\bigr|_{\mu_2} < 0$, and we find
$$
c_v (\mu_2 ) < 0 \ .
$$
For $\mu_c = \mu_3$, it follows from \secdQ\ that \eqn\tueo{ 0 <
Q''(\mu_3) = {1 \ov 2 b} - {1 \ov 4 \mu^2} - {1 \ov 4 \sqrt{\mu
-1} \mu^{3/2}} < {1 \ov 2b} } Plugging \tueo\ into \unhneC\ we
find:
$$
c_v (\mu_3) > 0 \ .
$$

\listrefs
\end
\end

Since $x \in [0,1)$, we can further expand it in a power series
 \eqn\adjpoly{
 S (x)
 = a x + b x^2 + c x^3 + \cdots  \ .
 }

As we indicated earlier the spirit of our discussion is similar to
that of the chiral lagrangian in QCD studies. The strategy has two
parts. First one can extract universal features from the effective
action. Secondly one can approximately determine the coefficients
of the effective action by explicitly matching, in our case, with
data in the dual supergravity description or the weakly coupled
theory. For example, Supergravity contains two known functions of
$t=T/R$, where $T$ is the temperature of the black hole and $R$ is
the radius of $AdS_5$. These are the Euclidean actions of the SBH
and BBH. We can in principle use this data to evaluate the
parameters of the phenomenological model for large values of the
't Hooft coupling.

Also note for $z \to -\infty$, we have \eqn\tuej{ \ZZ_2 \sim
|z|^{-{1 \ov 4}} e^{- {2i \ov 3} |z|^{3 \ov 2}} }

\appendix{B}{Details for solving \critone}

To analyze the solutions of \critone, it is convenient to rewrite
it in terms of an equation of $\rho_1$ defined by \whuo,
\eqn\critone{ a+2b \rt^2={1\over 4\rt (1-\rt)}, \qquad \ha \leq
\rho_1 \leq 1 } To solve \critone, we define the two functions:
$$
f_1(\rt)=a+2b\rt^2,
$$
$$
f_2(\rt)={1\over 4\rt (1-\rt)},
$$
both of which are convex\foot{We should point out that the
qualitative features of the solutions, like the number of
solutions and the existence of double solutions describing black
hole nucleation, are solely determined by the fact that
$f_1(\rho)$ is convex. This is also true for the range $\rho_1 <
1/2$}. The solutions to \critone\ correspond to
$f_1(\rt)=f_2(\rt)$. Since $f'_2(1/2)=0$ and $f'_1(1/2)= 2b>0$, we
have several possibilities:

\item{1.} $f_1(1/2)=a+b/2>1$ (the minimum value of $f_2=1$ happens
at $\rt=1/2$). Then there is a unique solution to \critone\ and
since the action goes to $\infty$ as $\rt\rightarrow 1$ the
extremum is a minimum with $Q''>0$.

\item{2.} $a+b/2=1$ (line III in fig.1). Since $b>0$ there are two
solutions, one at $\rt=1/2$, a local maximum, and a second with
$\rt>1/2$ which is a local minimum.

\item{3.} $a+b/2<1$, i.e. $f_1(1/2)<1$ there are several
possibilities. For $b$ small there is no solution, the curve $f_1$
stays always below $f_2$. This corresponds to the region between
the $a$-axis and line-I in fig.1. For a critical value of $b$ the
two curves will touch and their derivatives also coincide. This
defines line-I and implies:
$$
Q'=Q''=0
$$
There are double point solutions to $Q'=0$. Line-I satisfies the
simultaneous equations: \eqn\lineone{ a+2b\rt^2={1\over
4\rt(1-\rt)} \qquad a+6b\rt^2={1\over 4(1-\rt)^2} } Below this
line $Q$ of \egst\ has no critical points. A parametric
description of line-I follows from \lineone : \eqn\lineonepar{
a(\rho)={3\over 8\rho(1-\rho)}-{1\over 8(1-\rho)^2}}
$$
b(\rho)={1\over 16\rho^2 (1-\rho)^2}-{1\over 16\rho^3(1-\rho)}
$$

Between line-I and line-III ($a+b/2=1$) there are two critical
points for \egst, one close to $1/2$, the local maximum, and one
close to $1$, the local maximum. In between lines I-III, we have a
special line which will be associated with the HP-transition, and
it is defined by the set of critical points \critone for which the
action vanishes: \eqn\linetwo{Q'(\rt)=0, \qquad Q(\rt)=0} This is
line-II in fig.1. Hence below line II the absolute minimum for
$\ZZ$ corresponds to $\rt=0$, while above line-II we find that the
local minimum close to $\rt=1$ becomes the absolute maximum. We
can plot the behavior of the critical points and the value of the
action between lines I-III as follows. We parametrize this region
by first choosing a particular value of $\rho\in [1/2,1]$ in
\lineonepar\ and then draw a circle centered at $(a,b)=(1,0)$ with
radius $|a(\rho)-1+i b(\rho)|$ until it hits line-III, the angular
interval of course depends on $\rho$. In fig.2 we have plotted
along this arc for $\rho=.7$ the two values of the critical points
for the action. As expected they coincide in line I, which here is
the rightmost point on the curve, and then they separate, one
towards $1/2$ and the other towards $1$ as we approach line III.
In fig.4 we plot the critical values of $\rho$ and in fig.5 their
corresponding actions.

\ifig\ftwo{ We represent here the values of the two critical
points for \critone between lines I and III for a value of
$\rho=.7$ in \lineonepar. The horizontal axis represents the arc
of circle with center $(1,0)$ joining the two lines. We see that
in the rightmost part of the graph both eigenvalues coincide at
line I, and then they separate as we move towards line III, one
moves towards $1$ and the other towards $1/2$.} {\epsfxsize=8cm
\epsfbox{fig2.eps}}

\ifig\fthree{As in the previous figure, but now the upper and
lower curves represente respectively the values of the action at
the critical points.} {\epsfxsize=8cm \epsfbox{fig3.eps}}

\appendix{B}{Details for $(a,b)$ model}

In this appendix, we apply the general discussion of appendix A to
the $(a,b)$ model \themodel\ in slightly more detail. For
\themodel\ we have
 \eqn\trjp{
 S (x) = a x + b x^2
 }
and \Impote\ can be further written as
 $$
  V [\rho] = (1-a) |\rho_1|^2 - b |\rho_1|^4 + \sum_{n=2}^\infty {1 \ov n}
\, |\rho_n|^2
 $$
 The dependence on $\rho_1$ indicates that:
(i) when $b
0$, there should be a first order phase transition at some value
$a < 1$; (ii) when $b < 0$, there is a second order phase
transition at $a=1$. We restrict to our discussion to $b>0$.

For \themodel\ with $S(x)$ given by \trjp, \anpoP\ and the second
line of \seldcon\ becomes
 \eqn\phast{
 \rho_1^2 = {(1-a) \ov 2b}, \qquad 0
 \leq \rho_1^2 < {1 \ov 4}
 }
and
 \eqn\critonei{
  a+2b x={1\over 4\sqrt{x} (1-\sqrt{x})}, \qquad {1 \ov 4} \leq
 x \leq 1
 }
 We note that \phaso,\phast\ and \critonei\ are precisely \theads,
\theyo\ and \critone. The Jacobian in \patI\ is, however, not
known exactly. Thus it is inconvenient to use this method to
understand finite $N$ physics.

To analyze the solutions of \critone, it is convenient to rewrite
it in terms of an equation of $\rho_1$ defined by \whuo,
\eqn\critone{ a+2b \rt^2={1\over 4\rt (1-\rt)}, \qquad \ha \leq
\rho_1 \leq 1 } To solve \critone, we define the two functions:
$$
f_1(\rt)=a+2b\rt^2,
$$
$$
f_2(\rt)={1\over 4\rt (1-\rt)},
$$
both of which are convex\foot{We should point out that the
qualitative features of the solutions, like the number of
solutions and the existence of double solutions describing black
hole nucleation, are solely determined by the fact that
$f_1(\rho)$ is convex. This is also true for the range $\rho_1 <
1/2$}. The solutions to \critone\ correspond to
$f_1(\rt)=f_2(\rt)$. Since $f'_2(1/2)=0$ and $f'_1(1/2)= 2b>0$, we
have several possibilities:

\item{1.} $f_1(1/2)=a+b/2>1$ (the minimum value of $f_2=1$ happens
at $\rt=1/2$). Then there is a unique solution to \critone\ and
since the action goes to $\infty$ as $\rt\rightarrow 1$ the
extremum is a minimum with $Q''>0$.

\item{2.} $a+b/2=1$ (line III in fig.1). Since $b>0$ there are two
solutions, one at $\rt=1/2$, a local maximum, and a second with
$\rt>1/2$ which is a local minimum.

\item{3.} $a+b/2<1$, i.e. $f_1(1/2)<1$ there are several
possibilities. For $b$ small there is no solution, the curve $f_1$
stays always below $f_2$. This corresponds to the region between
the $a$-axis and line-I in fig.1. For a critical value of $b$ the
two curves will touch and their derivatives also coincide. This
defines line-I and implies:
$$
Q'=Q''=0
$$
There are double point solutions to $Q'=0$. Line-I satisfies the
simultaneous equations: \eqn\lineone{ a+2b\rt^2={1\over
4\rt(1-\rt)} \qquad a+6b\rt^2={1\over 4(1-\rt)^2} } Below this
line $Q$ of \egst\ has no critical points. A parametric
description of line-I follows from \lineone : \eqn\lineonepar{
a(\rho)={3\over 8\rho(1-\rho)}-{1\over 8(1-\rho)^2}}
$$
b(\rho)={1\over 16\rho^2 (1-\rho)^2}-{1\over 16\rho^3(1-\rho)}
$$

Between line-I and line-III ($a+b/2=1$) there are two critical
points for \egst, one close to $1/2$, the local maximum, and one
close to $1$, the local maximum. In between lines I-III, we have a
special line which will be associated with the HP-transition, and
it is defined by the set of critical points \critone for which the
action vanishes: \eqn\linetwo{Q'(\rt)=0, \qquad Q(\rt)=0} This is
line-II in fig.1. Hence below line II the absolute minimum for
$\ZZ$ corresponds to $\rt=0$, while above line-II we find that the
local minimum close to $\rt=1$ becomes the absolute maximum. We
can plot the behavior of the critical points and the value of the
action between lines I-III as follows. We parametrize this region
by first choosing a particular value of $\rho\in [1/2,1]$ in
\lineonepar\ and then draw a circle centered at $(a,b)=(1,0)$ with
radius $|a(\rho)-1+i b(\rho)|$ until it hits line-III, the angular
interval of course depends on $\rho$. In fig.2 we have plotted
along this arc for $\rho=.7$ the two values of the critical points
for the action. As expected they coincide in line I, which here is
the rightmost point on the curve, and then they separate, one
towards $1/2$ and the other towards $1$ as we approach line III.
In fig.4 we plot the critical values of $\rho$ and in fig.5 their
corresponding actions.

\ifig\ftwo{ We represent here the values of the two critical
points for \critone between lines I and III for a value of
$\rho=.7$ in \lineonepar. The horizontal axis represents the arc
of circle with center $(1,0)$ joining the two lines. We see that
in the rightmost part of the graph both eigenvalues coincide at
line I, and then they separate as we move towards line III, one
moves towards $1$ and the other towards $1/2$.} {\epsfxsize=8cm
\epsfbox{fig2.eps}}

\ifig\fthree{As in the previous figure, but now the upper and
lower curves represente respectively the values of the action at
the critical points.} {\epsfxsize=8cm \epsfbox{fig3.eps}}